\documentclass{pasj00}
\usepackage{graphicx}
\draft
\newcommand{\lw}[1]{\smash{\lower2.ex\hbox{#1}}}
\begin{document}
\SetRunningHead{Furuya et al.}{Dense Cores in GF\,9 Filament}
\Received{2007/11/20}%yyyy/mm/dd}
\Accepted{}%{yyyy/mm/dd}

\title{Low-Mass Star Forming Cores in the GF\,9 Filament
  \thanks{Accepted by PASJ (vol.\,60); preprint with the original quality 
figures are available at \texttt{
http://subarutelescope.org/staff/rsf/publication.html}
  }
}
%%% begin:list of authors
% Do NOT capitalize all letters in "textsc".
%\author{A-Firstname \textsc{A-Familyname} %
%  \thanks{Example: Present Address is xxxxxxxxxx}}
%\affil{A-Address of Institute}
%\email{aaaaa@xxx.xxx.xx.xx}
\author{Ray S. Furuya} %\altaffilmark{1}}
\affil{Subaru Telescope, National Astronomical Observatory of Japan,\\
650 North A'ohoku Place, Hilo, HI\,96720, U.S.A.}
\email{rsf@subaru.naoj.org}

\author{Yoshimi Kitamura} %\altaffilmark{2}}
\affil{Institute of Space and Astronautical Science, Japan Aerospace Exploration Agency,\\
3-1-1 Yoshinodai, Sagamihara, Kanagawa 229-8510}
\email{kitamura@isas.jaxa.jp}

\and

\author{Hiroko Shinnaga} %\altaffilmark{3}}
\affil{Caltech Submillimeter Observatory, California Institute of Technology,\\
111 Nowelo Street, Hilo, HI\,96720, U.S.A.}
\email{shinnaga@submm.caltech.edu}

%%% Please use the following style in case that sorting by 
%%% affilation is impossible. 
%
% \author{%
%   D-Firstname \textsc{D-Familyname}\altaffilmark{1}
%   E-Firstname \textsc{E-Familyname}\altaffilmark{1,2}
%   and
%   F-Firstname \textsc{F-Familyname}\altaffilmark{2}}
% \altaffiltext{1}{Address of Institute}
% \email{ddddd@xxx.xxx.xx.xx}
% \email{eeeee@xxx.xxx.xx.xx}
% \altaffiltext{2}{Address of Institute}


\maketitle

\begin{abstract}
We carried out an unbiased mapping survey of dense molecular cloud cores
traced by the NH$_3$ (1,1) and (2,2) inversion lines in the GF\,9 filament 
which contains an extremely young low-mass protostar GF\,9-2 
(Furuya et al. 2006, ApJ, 653, 1369).
The survey was conducted using the Nobeyama 45\,m telescope
over a region of $\sim 1.5^{\circ}\times 1^{\circ}$
with an angular resolution of \timeform{73"}.
The large-scale map revealed that the filament contains at least 7 dense cores, 
as well as 3 possible ones, located at regular intervals of $\sim 0.9$\,pc.
Our analysis shows that these cores have 
kinetic temperatures of $\lesssim$ 10\,K
and LTE-masses of 1.8 -- 8.2 \MO, 
making them typical sites of low-mass star formation.
All the identified cores are likely to be gravitationally unstable 
because their LTE-masses are larger than their virial masses.
Since the LTE-masses and separations of the cores are consistent with 
the Jeans masses and lengths, respectively, for the 
low-density ambient gas,
we argue that the identified cores have formed via the gravitational
fragmentation of the natal filamentary cloud.
\end{abstract}

%%%%%%%%%%%%%%%%%%%%%%%%%%%%%%%%%%%%%%%%%%%%%%%%%%%%%%%%%
\section{Introduction}
\label{s:intro}

A filamentary dark cloud often provides us with a unique opportunity to 
investigate the formation and evolution of a dense cloud core, 
the birthplace of low-mass stars, through fragmentation processes.
The dense cores appear to maintain themselves 
as substructures in the parental cloud that is known to be mostly governed
by supersonic turbulence.
The dense cores usually exhibit supersonic line widths; the turbulence
prevents the gravitational collapse of the cores.
It is now widely accepted that the 
turbulence plays a fundamental role in controlling core formation processes 
(e.g., Mac Low \& Klessen 2004; Ballesteros-Paredes et al. 2007 for reviews
and references therein).
Possible mechanisms for the core formation and collapse
have a widespread range from the gradual release of magnetic support 
(e.g., Shu, Adams \& Lizano 1987)
to the dynamical dissipation of turbulent waves 
(e.g., Larson 1981; Ostriker et al. 2001; Padoan \& Nordlund 2002).
Although we now have a good overall picture of the theory of low-mass star formation
(e.g., McKee \& Ostriker 2007), 
observational verification of how dense cores form and collapse
has produced limited success.\par

There have been quite a few observational studies which have
assessed the initial conditions for the gravitational 
collapse of a dense core. 
The low-mass protostar GF\,9-2 with bolometric luminosity and temperature of
$\simeq 0.3$\,\LO\  and $\leq 20$\,K, respectively (Wiesemeyer 1997),
remains the best-characterized extremely young protostar
that would be a missing link between starless
cores and class 0 protostars.
The natal core of GF\,9-2 is believed to retain the initial conditions of
the gravitational collapse because 
the central protostar has not launched an extensive outflow.
Namely the core has not yet been destroyed by an extensive outflow
(Furuya et al. 2006; hereafter paper I).
Our detailed analysis suggested that 
the core has undergone its gravitational
collapse for $\sim 2\times 10^5$ years (the free-fall time) from 
initially unstable state (Larson 1969; Penston 1969; Hunter 1977), 
and that the protostar has formed $\lesssimeq 5\times 10^3$ years ago.
Notice that the GF\,9-2 core is cross-identified as 
L\,1082C in Benson \& Myers (1989),
LM\,351 in Lee \& Myers (1999, hereafter LM99), and
GF\,9-Core in Ciardi et al. (2000); the central Young Stellar Object (YSO) is
also recognized as PSC\,20503$+$6006.\par

%%%%%%%%%%%%%%%%%%%%%%%%%%%%%%%%%%%%%%%%%%%%%%%%%%%%%%%%%%
Besides the presence of GF\,9-2, 
the filamentary dark cloud GF\,9 (Schneider \& Elmegreen 1979) would be an ideal laboratory
to establish a core formation scenario through the fragmentation of a filamentary cloud, 
because an ISO imaging survey in the far-infrared has demonstrated the presence of 
several class 0 and I sources (Wiesemeyer 1997; Wiesemeyer et al 1999).
Towards our ultimate goal of defining an observational framework for the cloud core formation,
the first step is to investigate the physical properties of dense cores formed in the filament.
We therefore performed an unbiased survey of dense cores in the GF\,9 filament.
The filament has an extent of $\sim 1.5^{\circ}\times 1.0^{\circ}$ 
in the optical image (Schneider \& Elmegreen 1979);
the eastern part of the cloud was firstly identified by Lynds (1962) as L\,1082.
The distance ($d$) to the GF\,9 cloud is still controversial,
as discussed in  Grenier et al.(1989) and
Poidevin \& Bastien (2006, and references therein).
In order to keep consistency with paper I, 
we adopt $d=$ 200\,pc, as reported by Wiesemeyer (1997), 
who derived the distance
towards GF\,9-2 based on star counts, instead of 440\,pc (Viotti 1969).

%%%%%%%%%%%%%%%%%%%%%%%%%%%%%%%%%%%%%%%%%%%%%%%%%%%%%%%%%
\section{Observations and Data Reduction}
\label{s:obs}

We carried out simultaneous observations of the NH$_3$ (1,1) and (2,2) lines
using the Nobeyama Radio Observatory (NRO)\footnote{Nobeyama Radio Observatory 
is a branch of the National Astronomical Observatory of Japan, 
National Institutes of Natural Sciences.} 45\,m telescope 
over 16 days in 2006 April.
We used the 22\,GHz cooled HEMT receiver (H22) which receives right- and left-hand
circular polarization components simultaneously.
The beam width ($\theta_{\rm HPBW}$) and main-beam efficiency ($\eta_{\rm mb}$)
of the telescope were \timeform{73"}$\pm$\timeform{0."2} 
and $83\pm 5$\,\%, respectively, at 23.0\,GHz.
To obtain the dual polarization data of the (1,1) line, 
we configured two Auto Correlator (AC)
spectrometers having 8\,MHz bandwidth with 1024 channels;
the newly enhanced capability provided us with a factor of
3 times higher velocity resolution than the previous one.
After on-line smoothing with the Hamming window function, 
the effective velocity resolution (\mbox{$\Delta v_{\rm res}$}) 
for the NH$_3$ (1,1) lines is 0.180\,km~s$^{-1}$.
For the (2,2) transition, we used two acousto-optical spectrometers (AOSs)
which provide \mbox{$\Delta v_{\rm res}$} of 0.494 km~s$^{-1}$.
Here, we adopt the rest frequencies of 23694.4955\,MHz for the (1,1) transition
(Ungerechts, Walmsley \& Winnewisser 1980; hereafter UWW80) and
23722.6336\,MHz for (2,2) (Lovas 1992) transition.\par

We performed mapping observations under full-beam sampling 
to cover the whole of the GF\,9 filament previously 
imaged with the H$_2$CO $1_{01}-1_{11}$ absorption line
(G{\"u}sten 1994). Using position-switching mode, 
we observed a total of 766 points by dividing the filament into 8 rectangular regions; 
each region was mapped with a grid spacing of \timeform{80"}.
The telescope pointing was checked every 4 hours and was found 
to be accurate within \timeform{5"}.
The daily variation of the H22 receiver gain was checked by the peak 
antenna temperature (\mbox{$T_{\rm A}^{\ast}$}) of the NH$_3$ (1,1) 
emission towards the GF\,9-2 core center.
We estimate that the final uncertainty in flux calibration is 22\%.
All the spectra were calibrated by the standard chopper wheel method
and converted into the main-beam brightness temperature 
(\mbox{$T_{\rm mb}$} $=~T_{\rm A}^*/\eta_{\rm mb}$) scale.
At the final stage of the spectral data reduction, 
the dual polarization data were concatenated to increase the signal-to-noise ratio (S/N).
Subsequently we made a total integrated intensity map with an effective 
resolution of $\theta_{\rm eff} =$\,\timeform{100"} from smoothed velocity channel maps 
by a Gaussian function with $\theta_{\rm HPBW} =$\,\timeform{68"}.\par

In addition, 
we performed deep single-point integrations
towards the approximate center positions of 9 cores on the basis of preliminary 
NH$_3$ maps made during the mapping observations.
These deep integrations were intended to obtain a better estimate of
the intensity ratio of the (1,1) to (2,2) transition,
and gave a 4--5 times lower noise level 
than that in the mapping observations.

%%%%%%%%%%%%%%%%%%%%%%%%%%%%%%%%%%%%%%%%%%%%%%%%%%%%%%%%%
\section{Results and Analysis}
\label{ss:ResAna}
%%%%%%%%%%%%%%%%%%%%%%%%%%%%%%%%%%%%%%%%%%%%%%%%%%%%%%%%%

In this section, we present the total integrated intensity map 
of the NH$_3$ (1,1) emission,
as well as the (1,1) and (2,2) spectra obtained through the deep integrations. 
Since the (2,2) emission was detected only towards the GF\,9-2 and -9 core centers, 
we do not present the map of the transition.

%%%%%%%%%%%%%%%%%%%%%%%%%%%%%%%%%%%%%%%%%%%%%%%%%%%%%%
\subsection{Identification of Dense Cloud Cores in the Filament} 
\label{ss:totmap}

\begin{figure}
\includegraphics[width=12.6cm,angle=0]{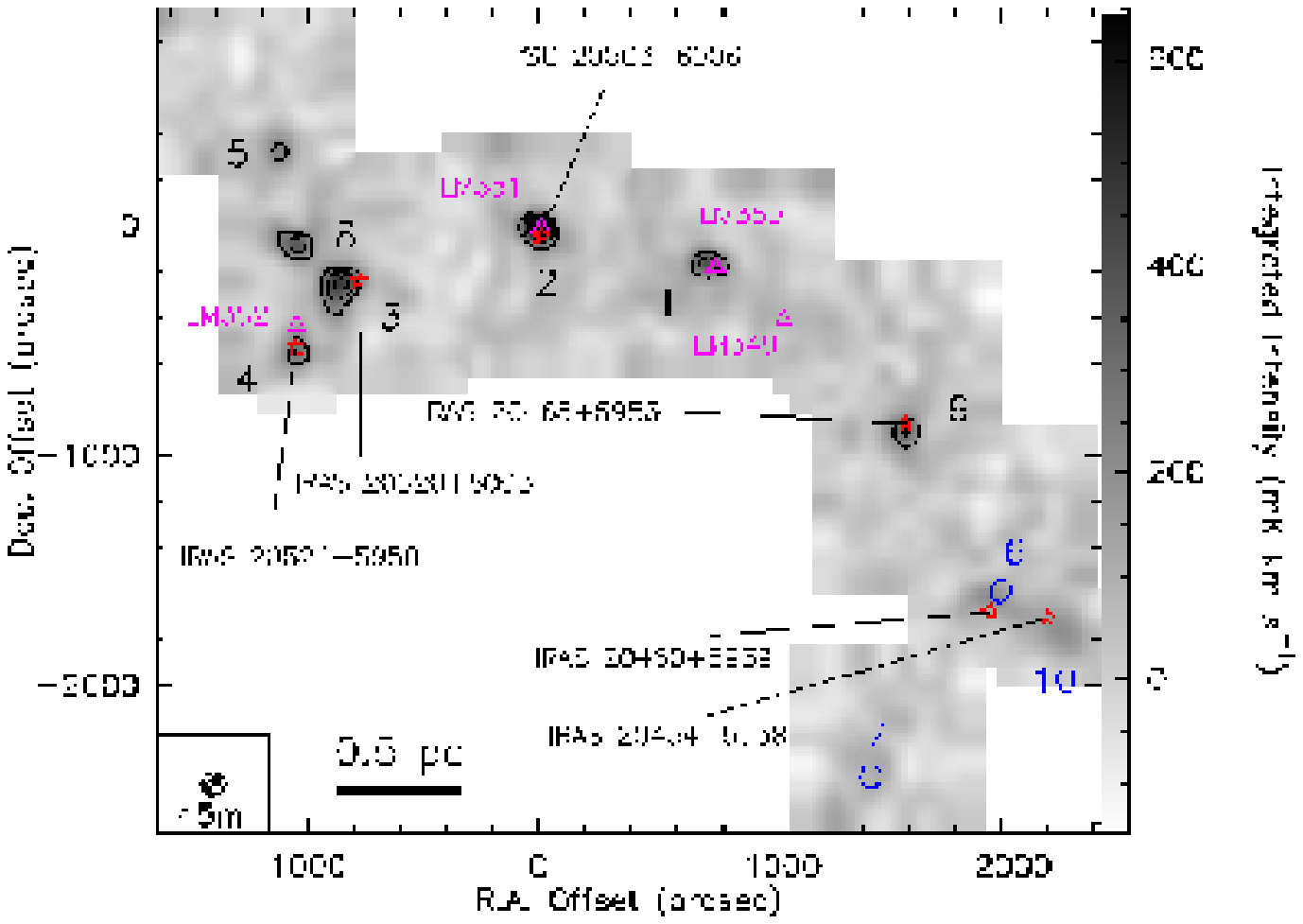}
\caption{Total integrated intensity map of the NH$_3$ (1,1) emission observed with the
Nobeyama 45\,m telescope. 
We integrated the main group of the hyperfine emission over the velocity range
$-3.2 \leq$ $V_{\rm LSR}$ /km~s$^{-1}$ $\leq -1.4$ (see section \ref{ss:totmap} for details).
The contour intervals are the 3$\sigma$ levels, starting from the 6$\sigma$
level where the 1$\sigma$ noise level is 37.0 mK km~s$^{-1}$ in \mbox{$T_{\rm mb}$}.
The map origin is R.A. $=$\,\timeform{20h51m29s.5},
Dec.$=$\,\timeform{60D18'38.0''} in J2000.0.
The peak positions of the identified cores and their identifications by
previous work are summarized in table \ref{tbl:cores}.
The thick black labels indicate the designation numbers of the cores, while
the blue thin labels the candidate cores.
The open blue circles indicate the areas to which the 45\,m telescope beam was pointed for
the deep integration (section \ref{s:obs}). 
Note that the size of the open blue circles is the same as
the effective beam size of the map ($\theta_{\rm eff} =\,$\timeform{100"},
see section \ref{s:obs}) that
is shown by the hatched circle in the bottom-left corner.
The open purple triangles and red stars with the labels indicate the positions of the
``embedded YSOs'' identified in Lee \& Myers (1999) and the IRAS sources,
respectively.
%%%%%%%
\label{fig:map}}
\end{figure}

\begin{figure}
\includegraphics[width=4.0cm,angle=0]{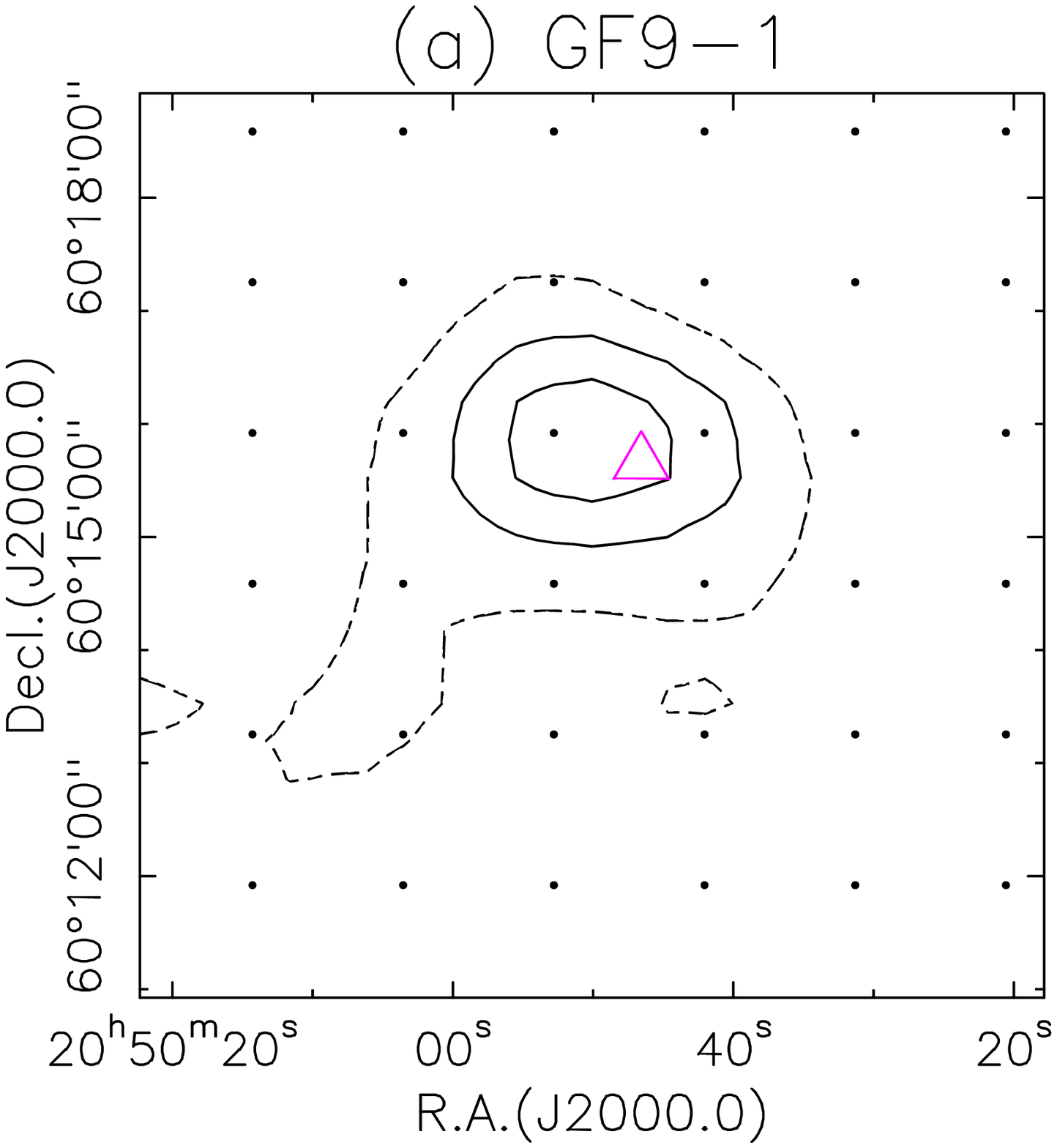}
\includegraphics[width=4.0cm,angle=0]{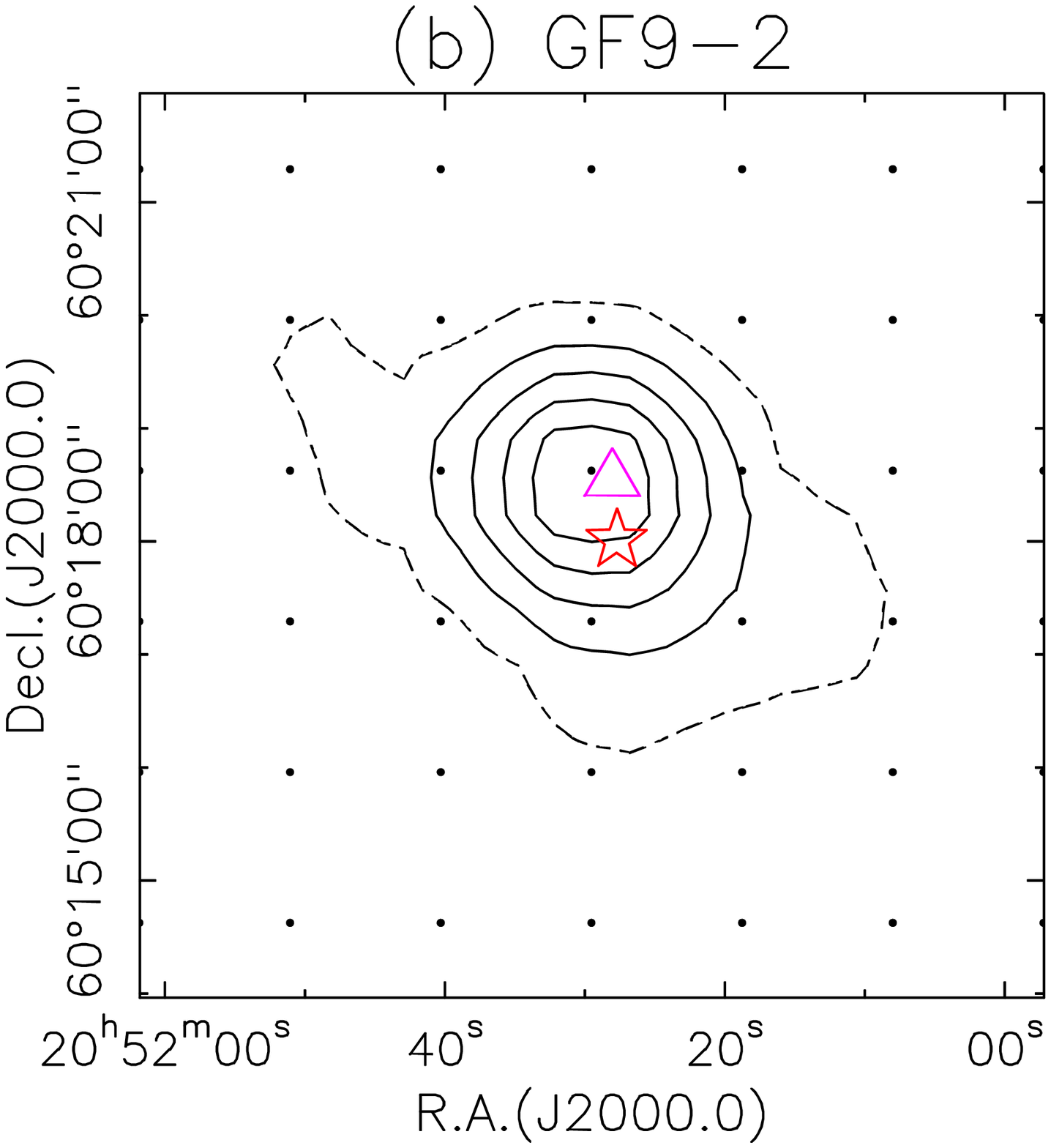}
\includegraphics[width=4.0cm,angle=0]{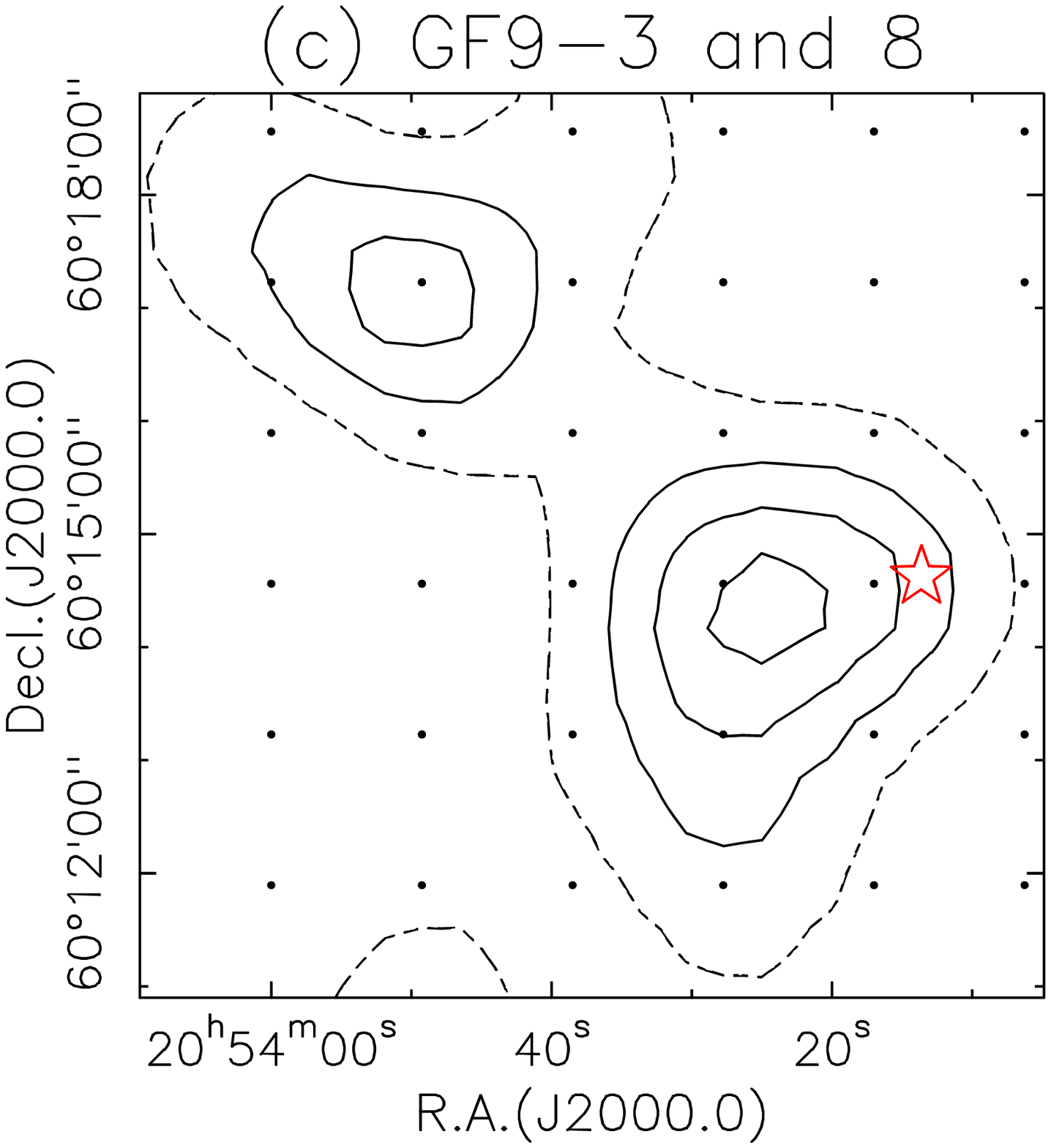} 
\includegraphics[width=4.1cm,angle=0]{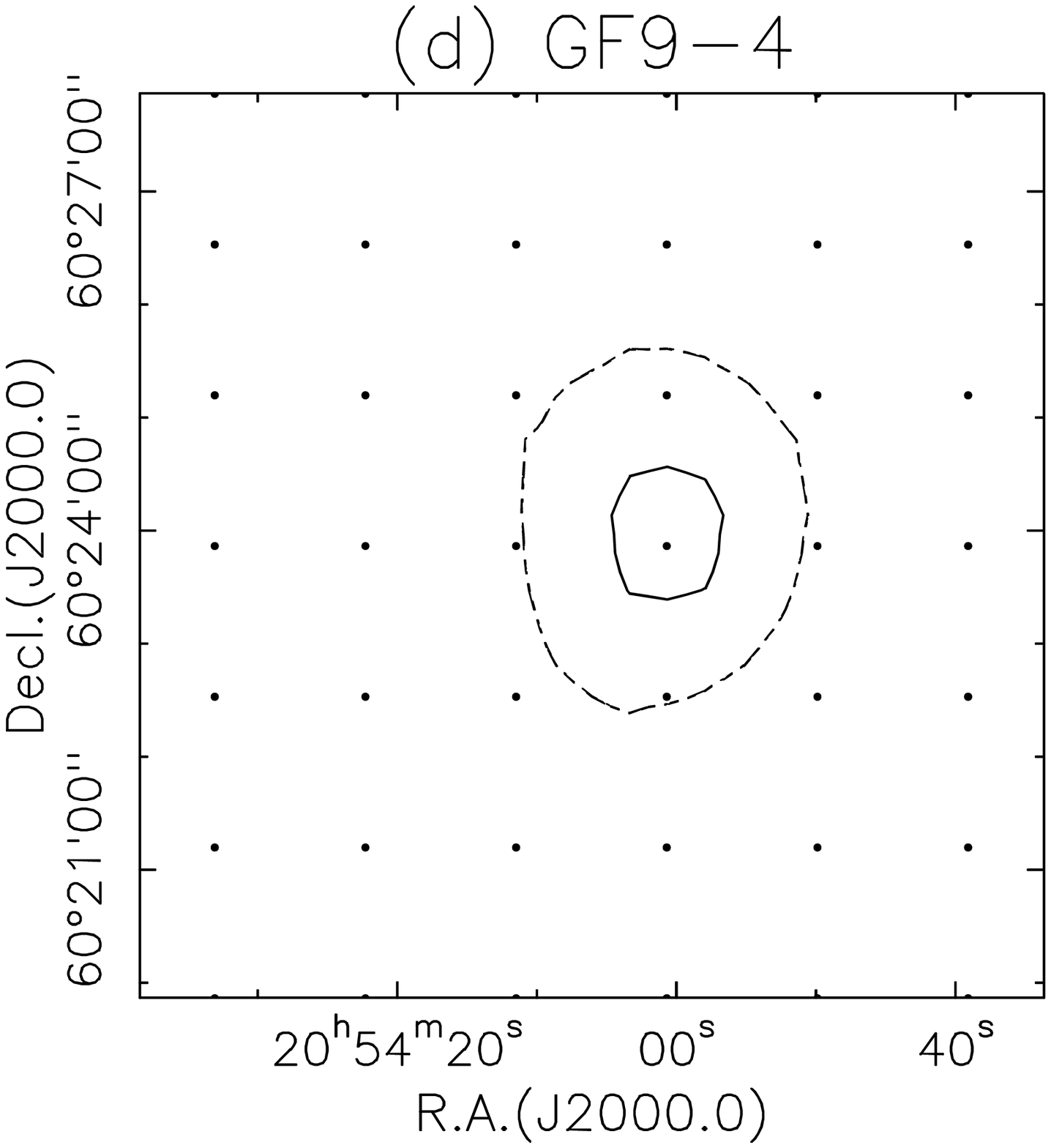} \\
\includegraphics[width=4.0cm,angle=0]{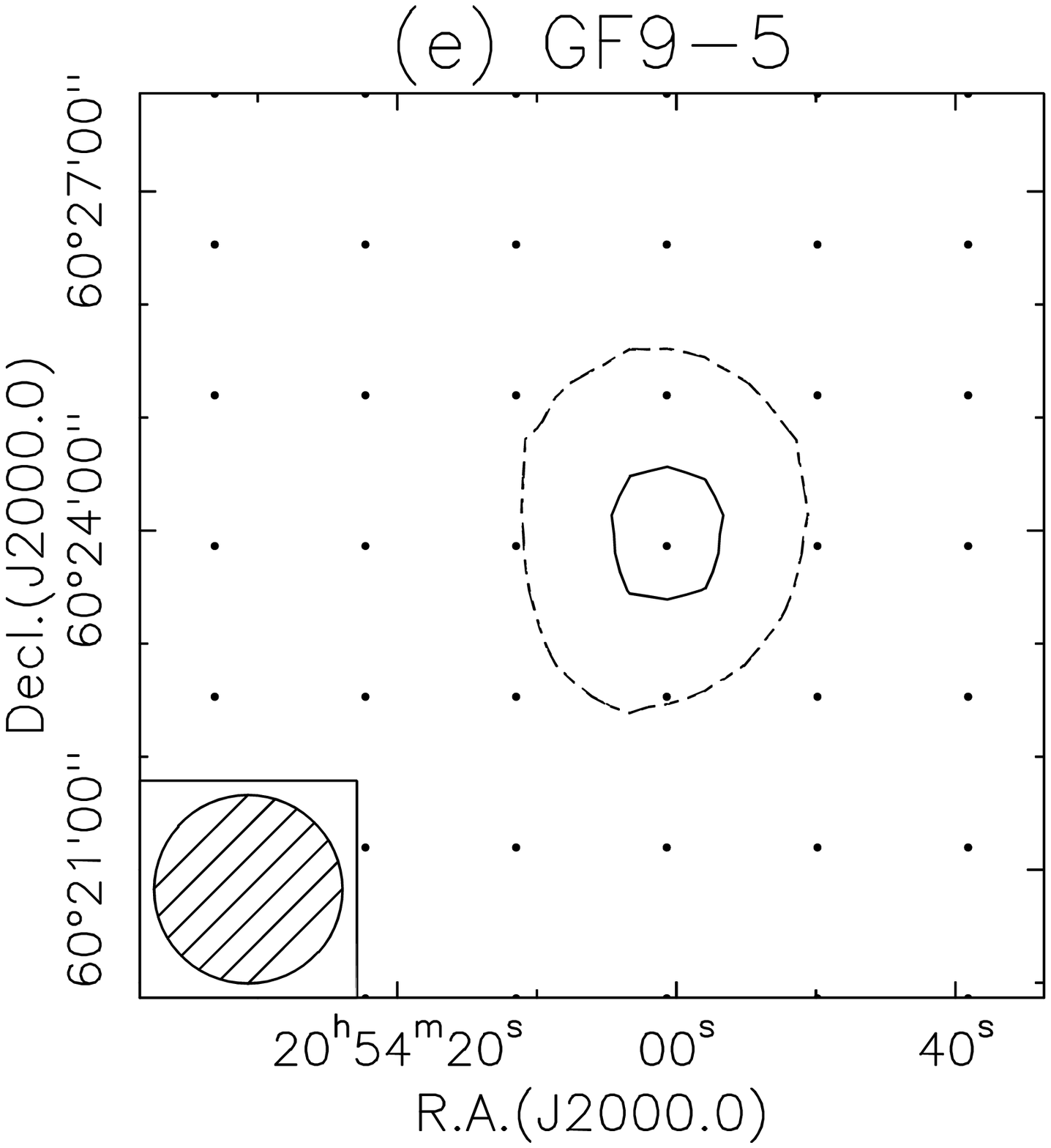}
\includegraphics[width=4.0cm,angle=0]{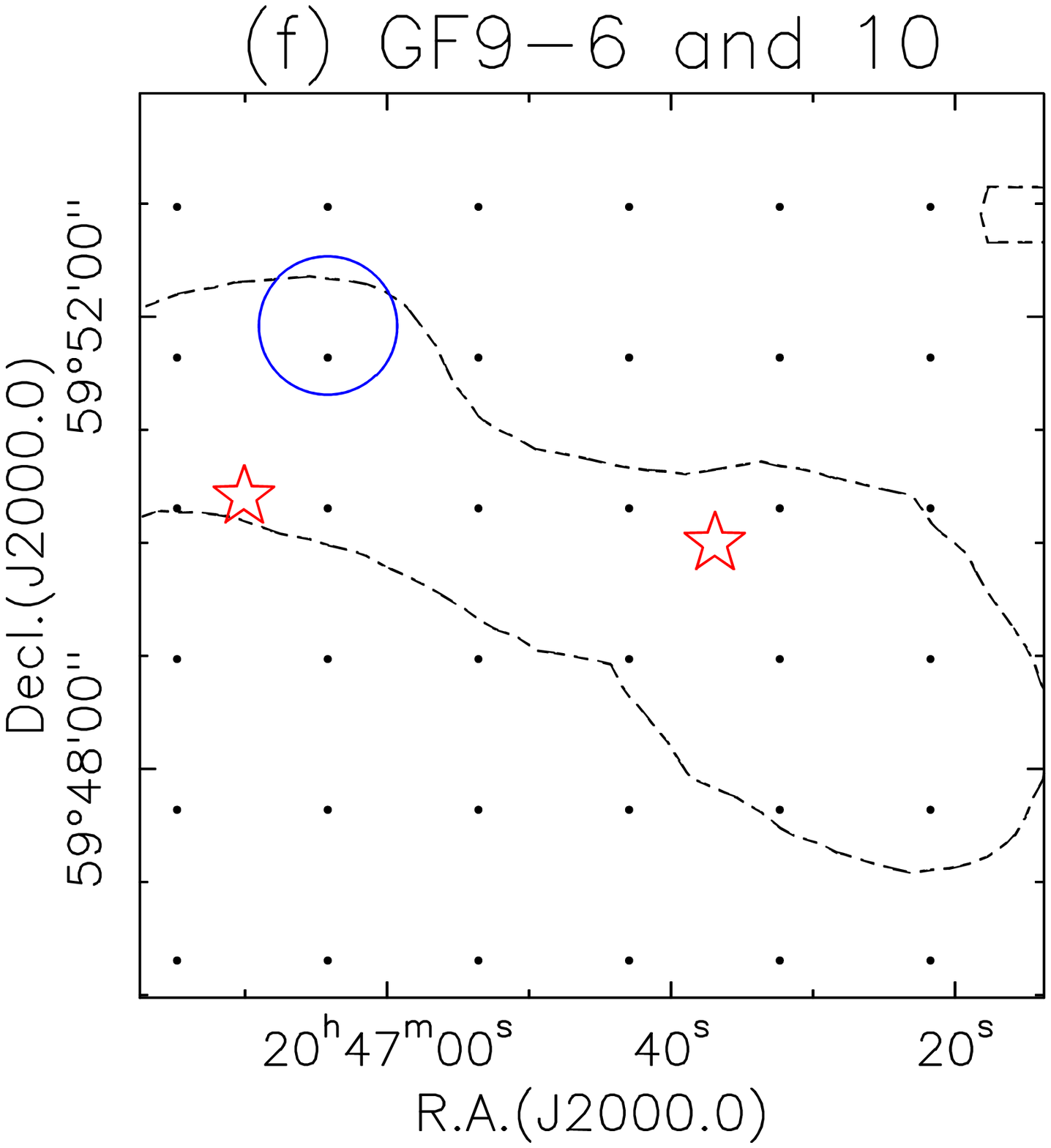} 
\includegraphics[width=4.0cm,angle=0]{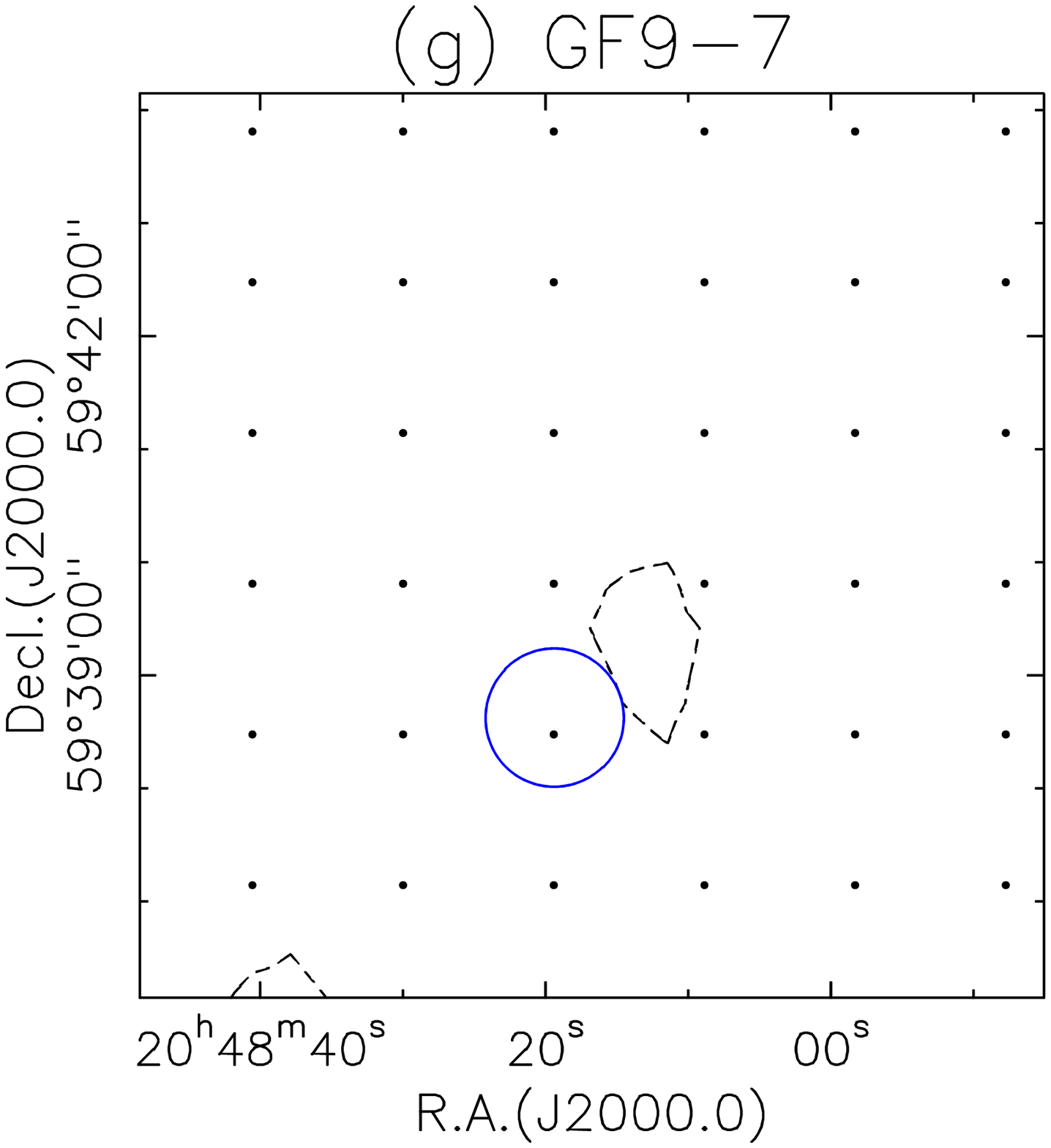}
\includegraphics[width=4.0cm,angle=0]{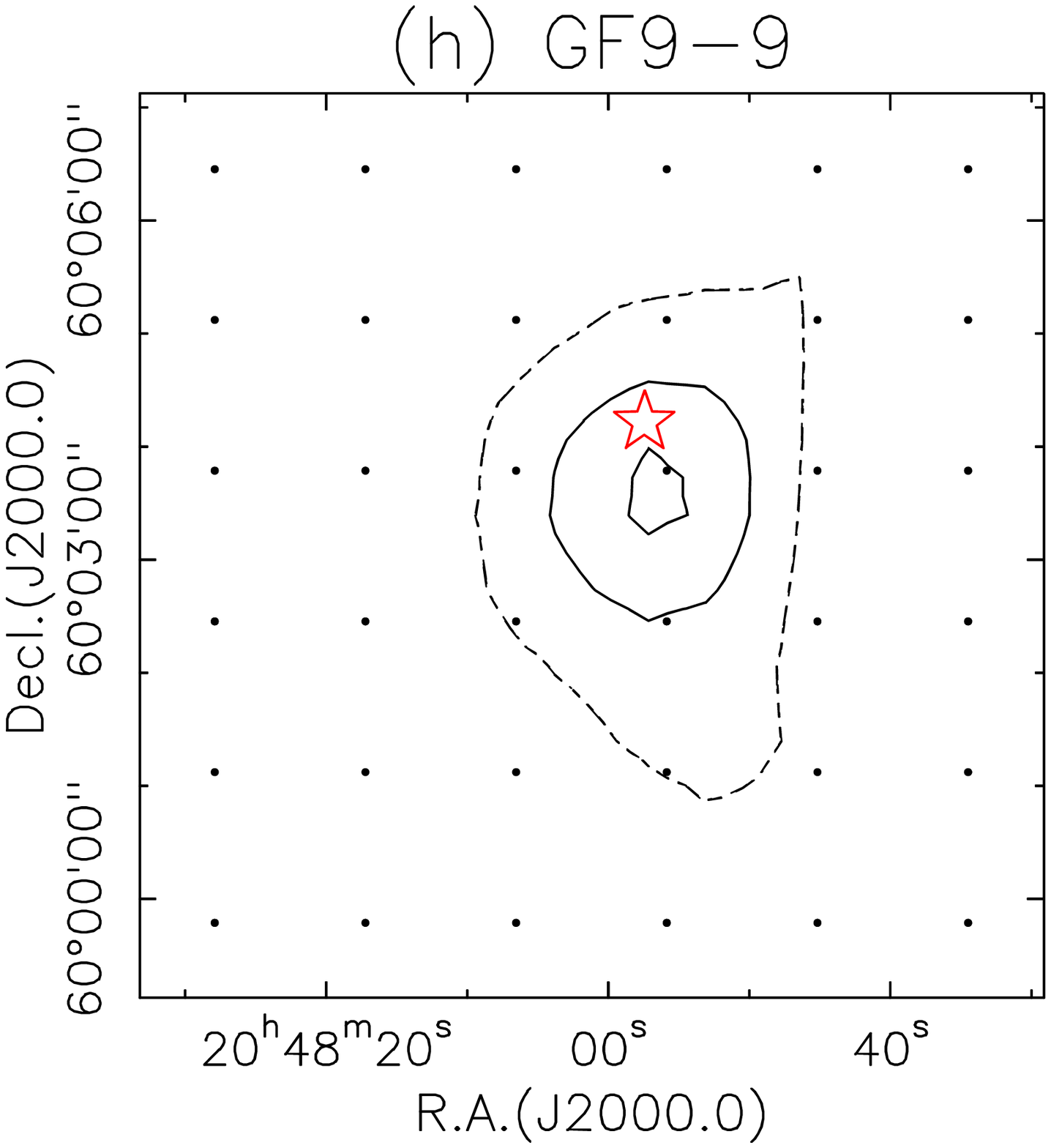}
\caption{Magnification of the NH$_3$ (1,1) total intensity map (figure \ref{fig:map}) 
towards the identified cores and the candidates (see table \ref{tbl:cores}).
Each panel has the same size of \timeform{6'.0} $\times$ \timeform{6'.0}.
The contour levels are the same as in figure \ref{fig:map}.
The dashed contours indicate the 3$\sigma$ level.
The small dots with a regular grid spacing of \timeform{80"} show the
observed grid points.
The hatched circle in the bottom left corner of panel (e) 
indicates the effective beam size of $\theta_{\rm eff} =$\,\timeform{100"}.
The other symbols are the same as those in figure \ref{fig:map}.
\label{fig:eachmap}}
\end{figure}

\begin{figure}
\includegraphics[width=12cm,angle=0]{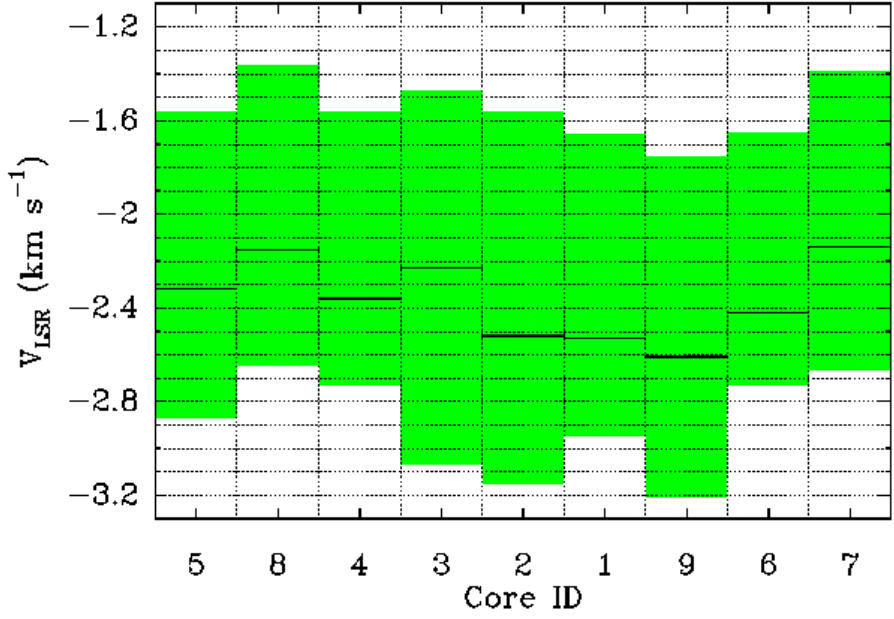}
\caption{LSR-velocity ranges of the main group of the NH$_3$ (1,1) hyperfine 
(HF) emission for the 9 spectra obtained through the deep single-pointing 
integration (section \ref{s:obs}).
The ranges are shown in the order of R.A. offset, except for GF\,9-7.
The horizontal thick bar in each bin
indicates the best-fit LSR-velocity, represented by $v_0$ in table \ref{tbl:sp}, 
from the hyperfine structure analysis (section \ref{ss:Ncol}). 
\label{fig:vrange}}
\end{figure}

\begin{figure}
\includegraphics[width=6.2cm,angle=0]{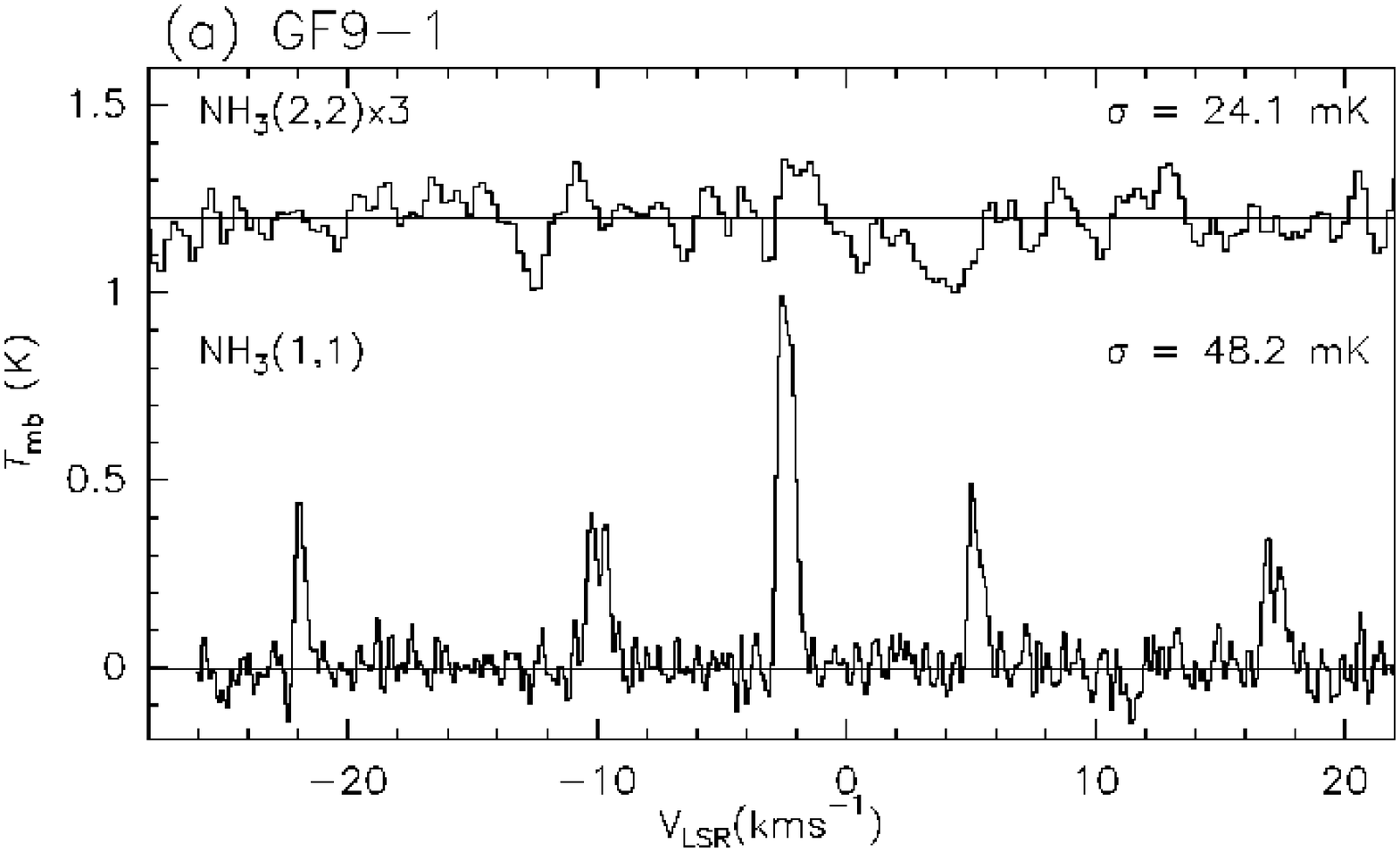}
\includegraphics[width=6.2cm,angle=0]{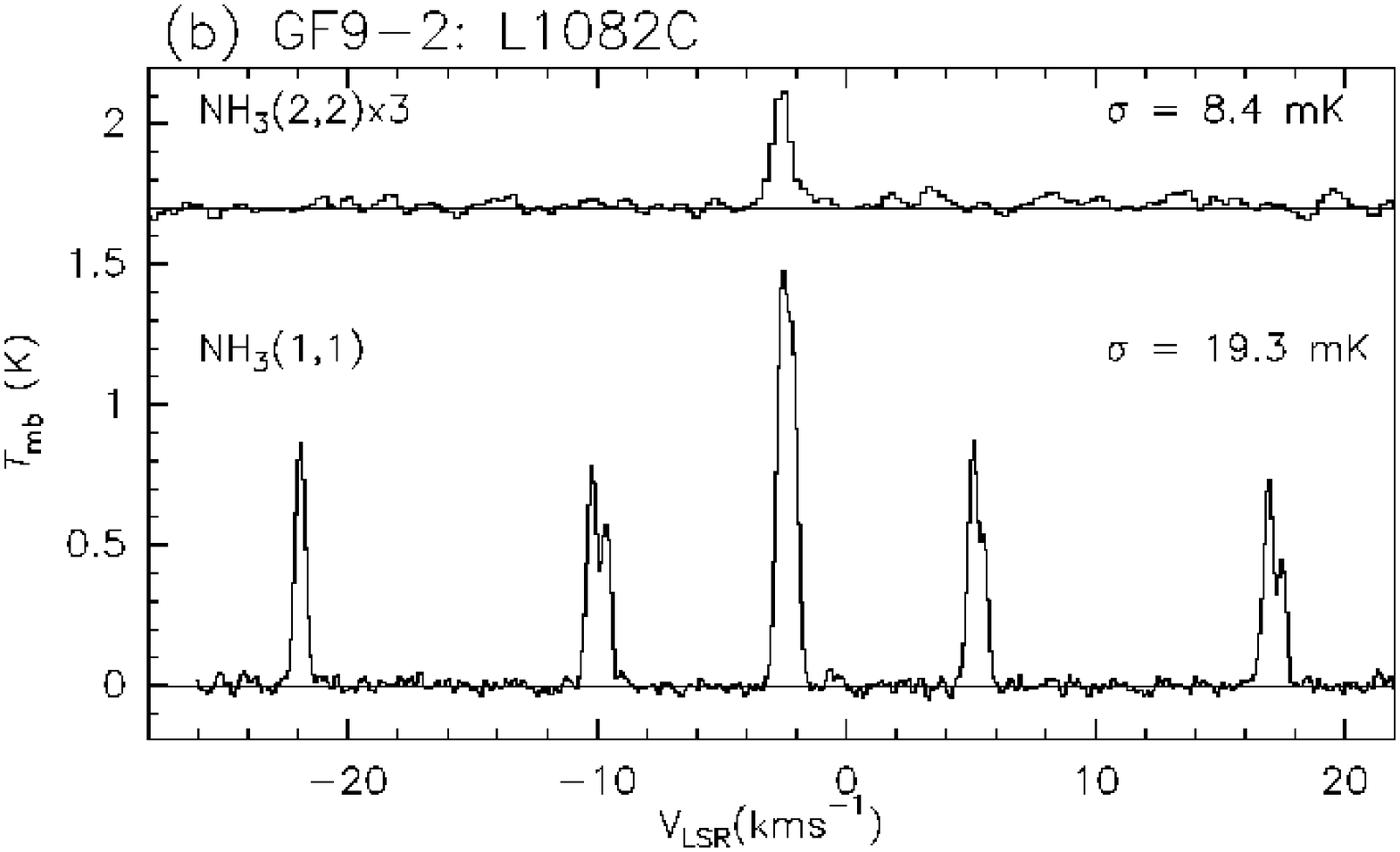} 
\includegraphics[width=6.2cm,angle=0]{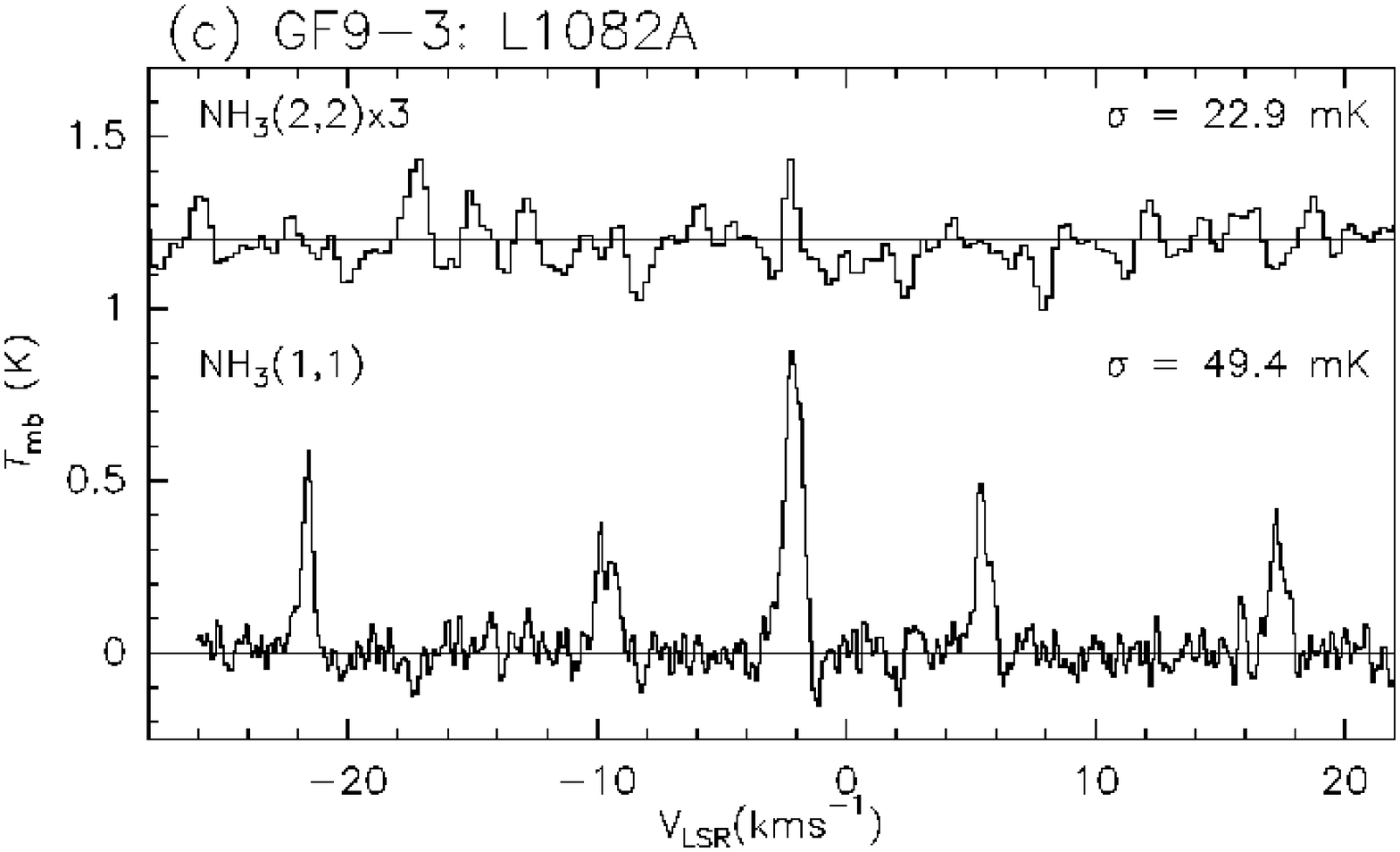}
\includegraphics[width=6.2cm,angle=0]{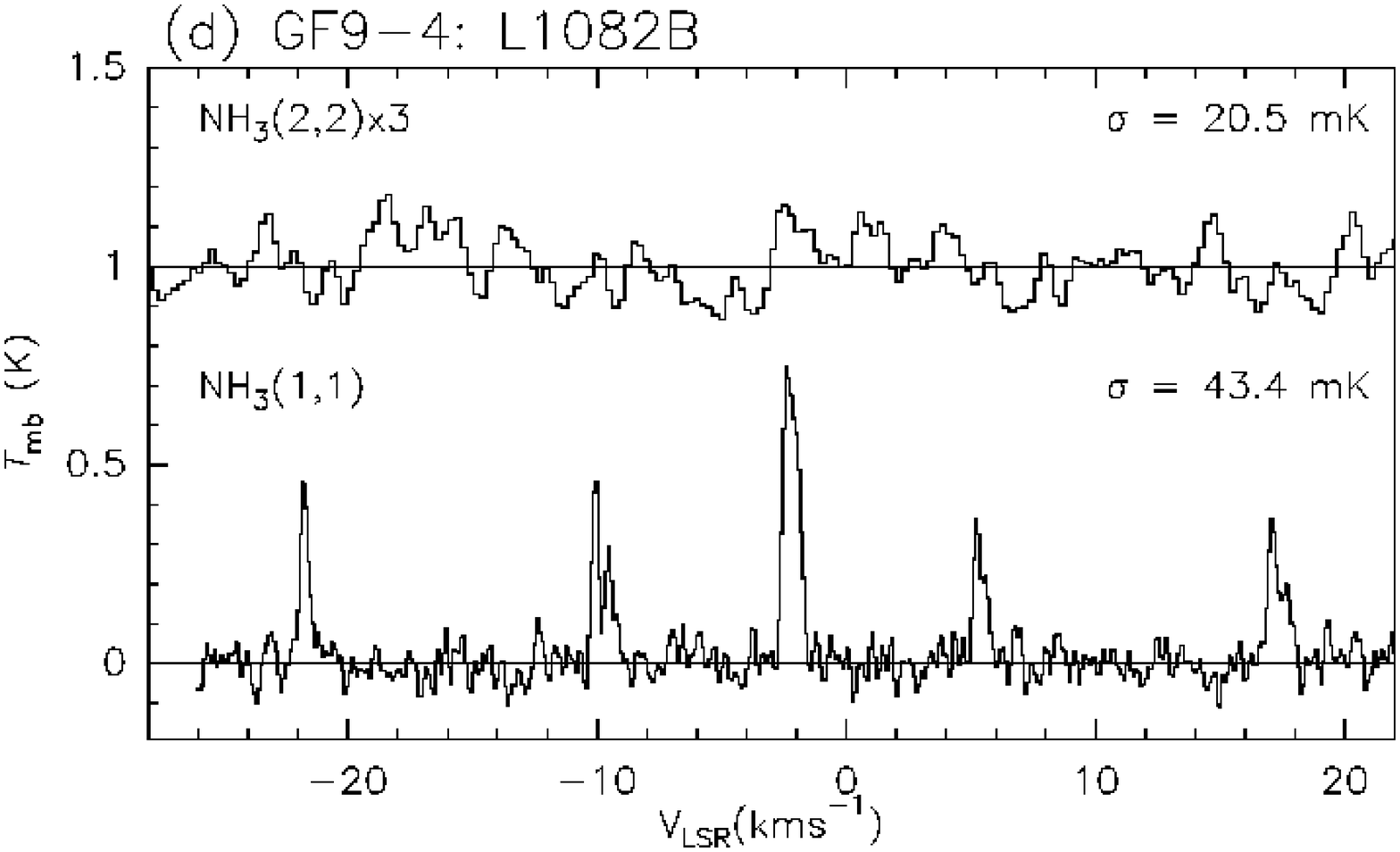} 
\includegraphics[width=6.2cm,angle=0]{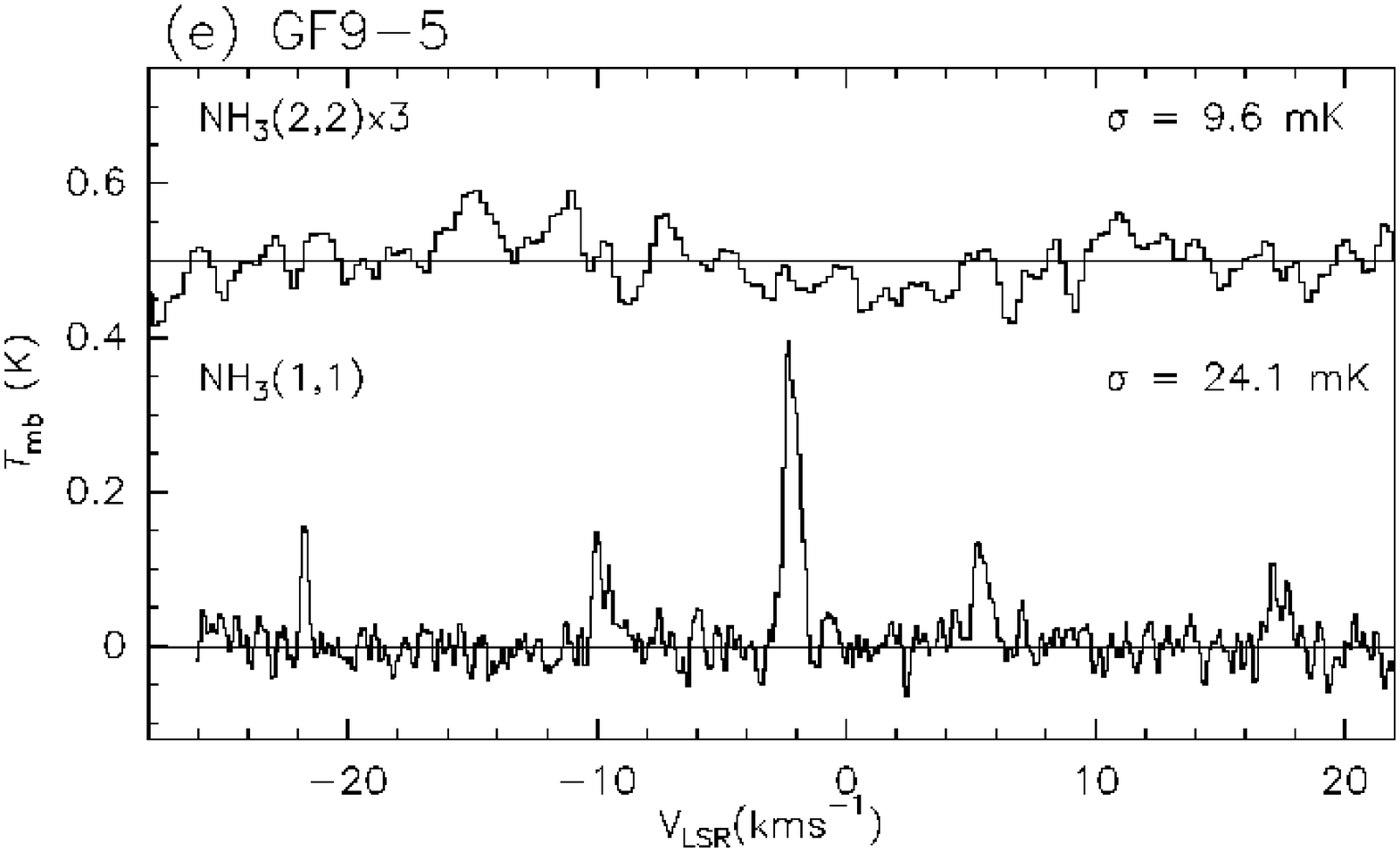}
\includegraphics[width=6.2cm,angle=0]{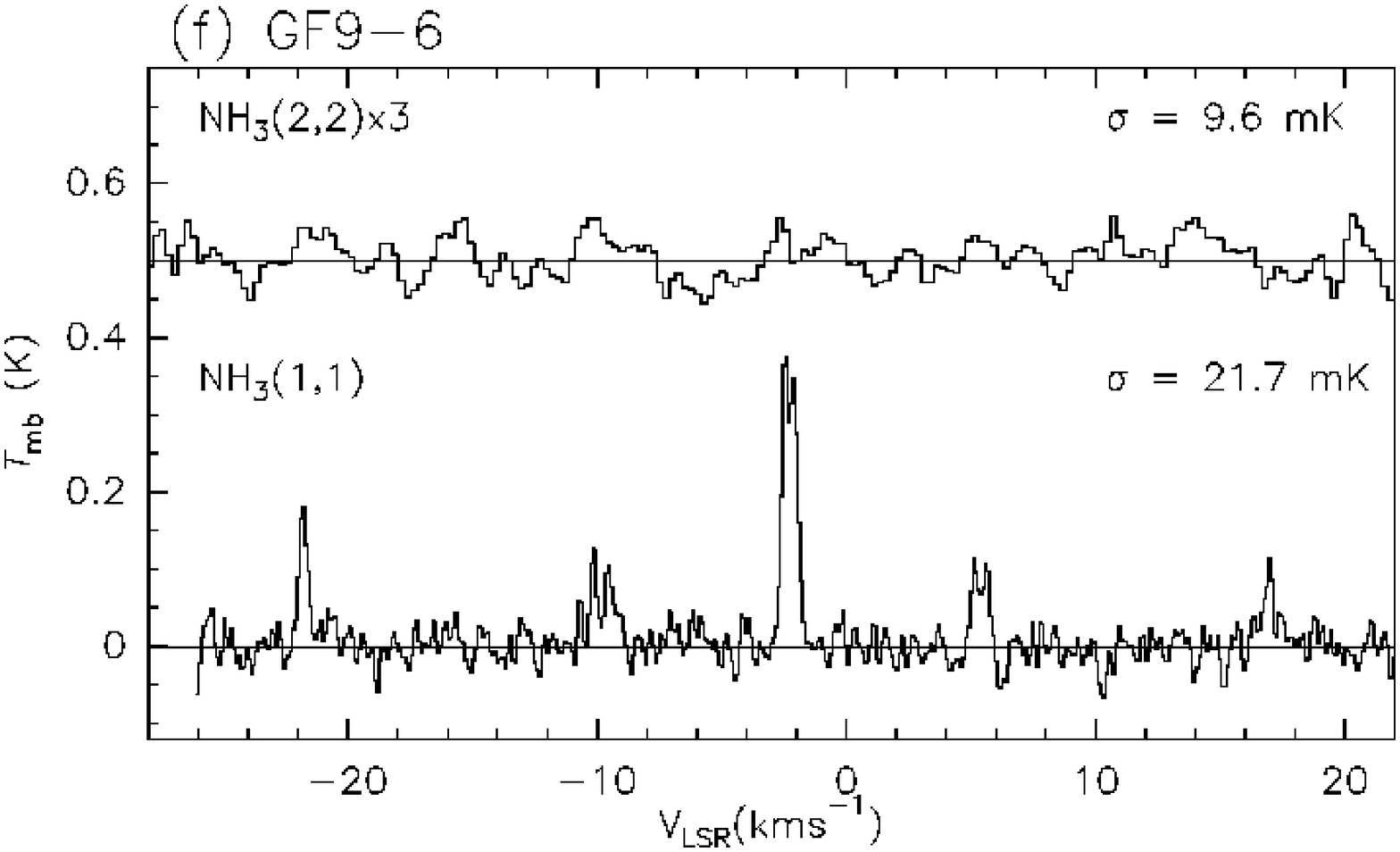} 
\includegraphics[width=6.2cm,angle=0]{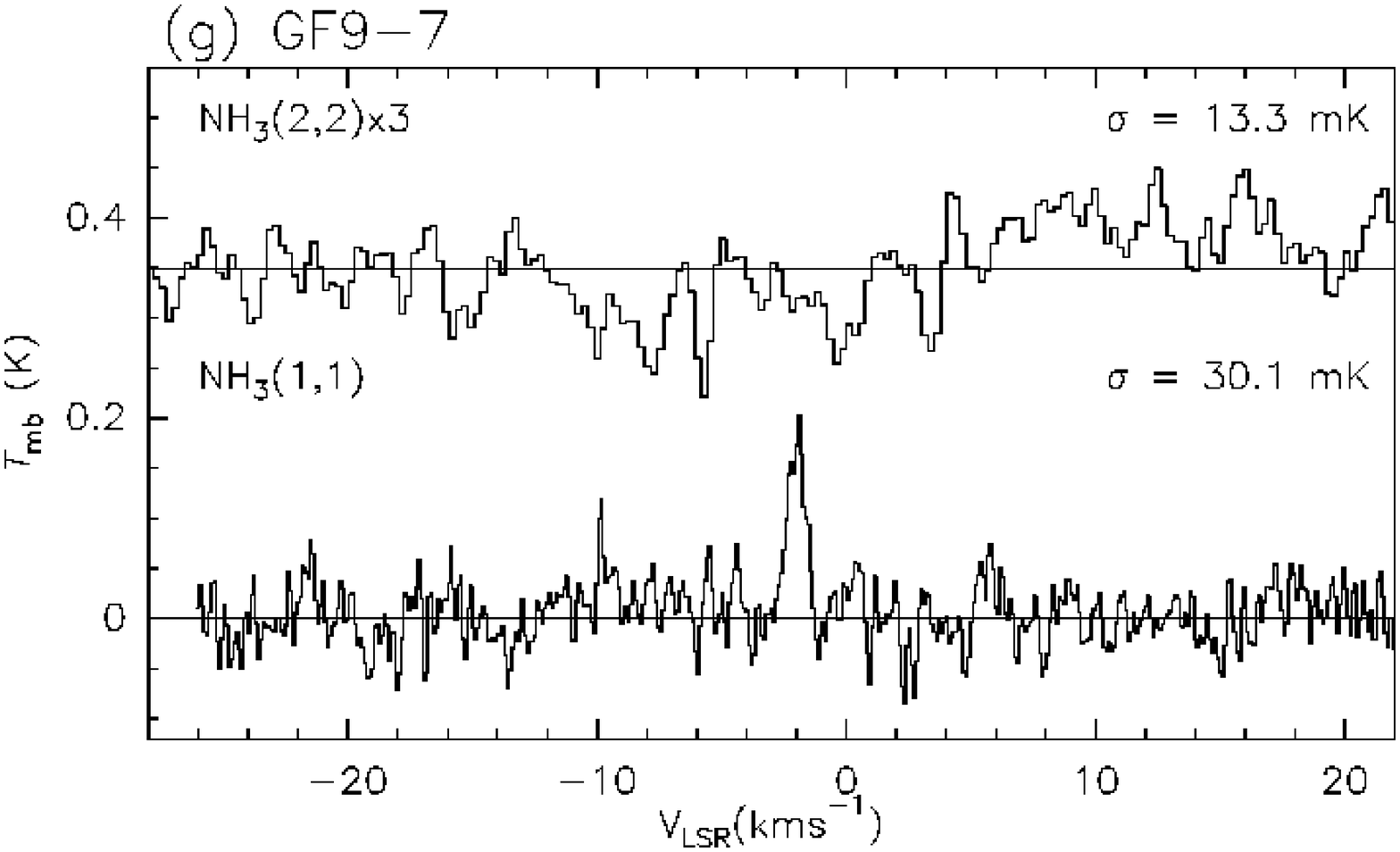}
\includegraphics[width=6.2cm,angle=0]{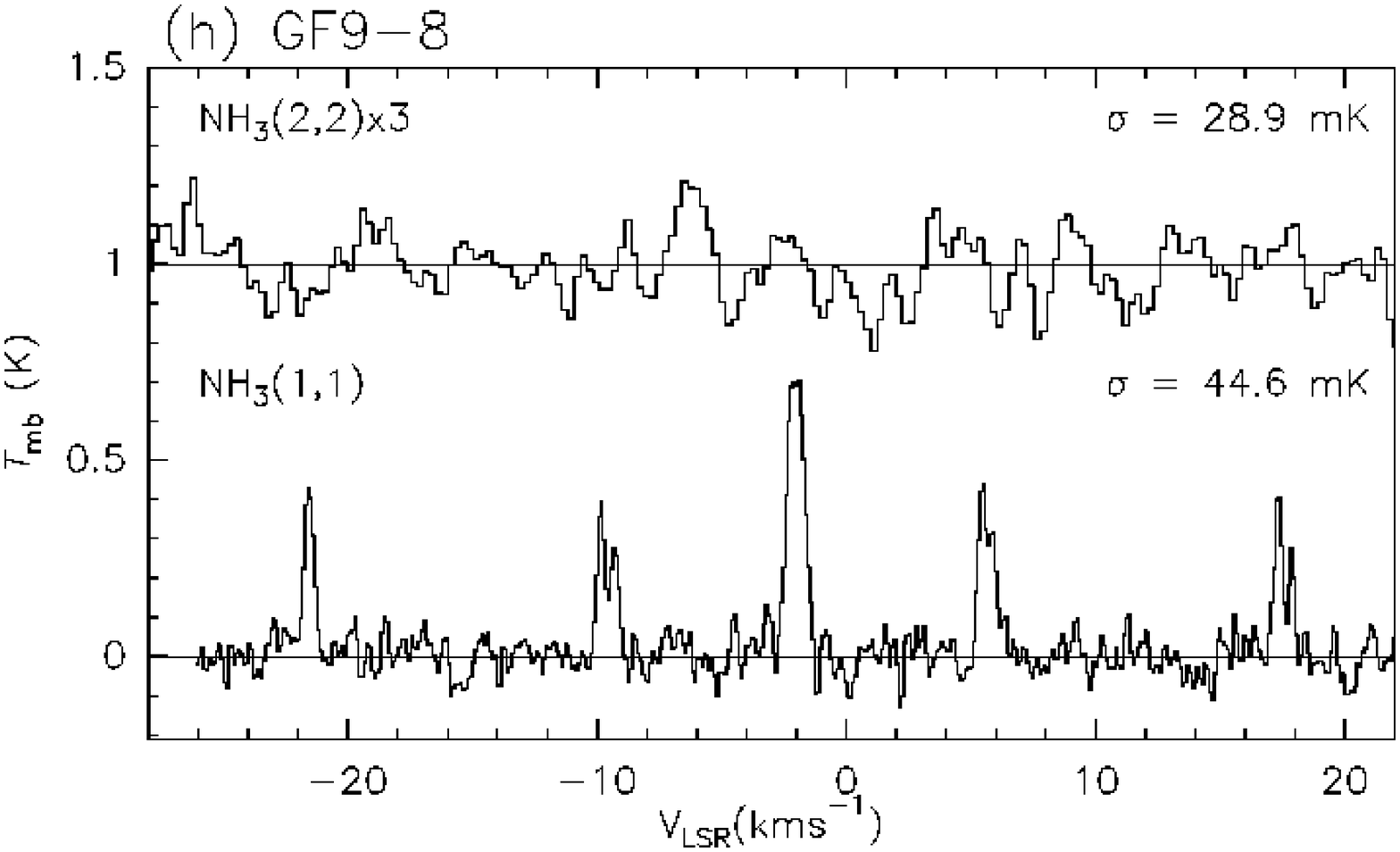} 
\includegraphics[width=6.2cm,angle=0]{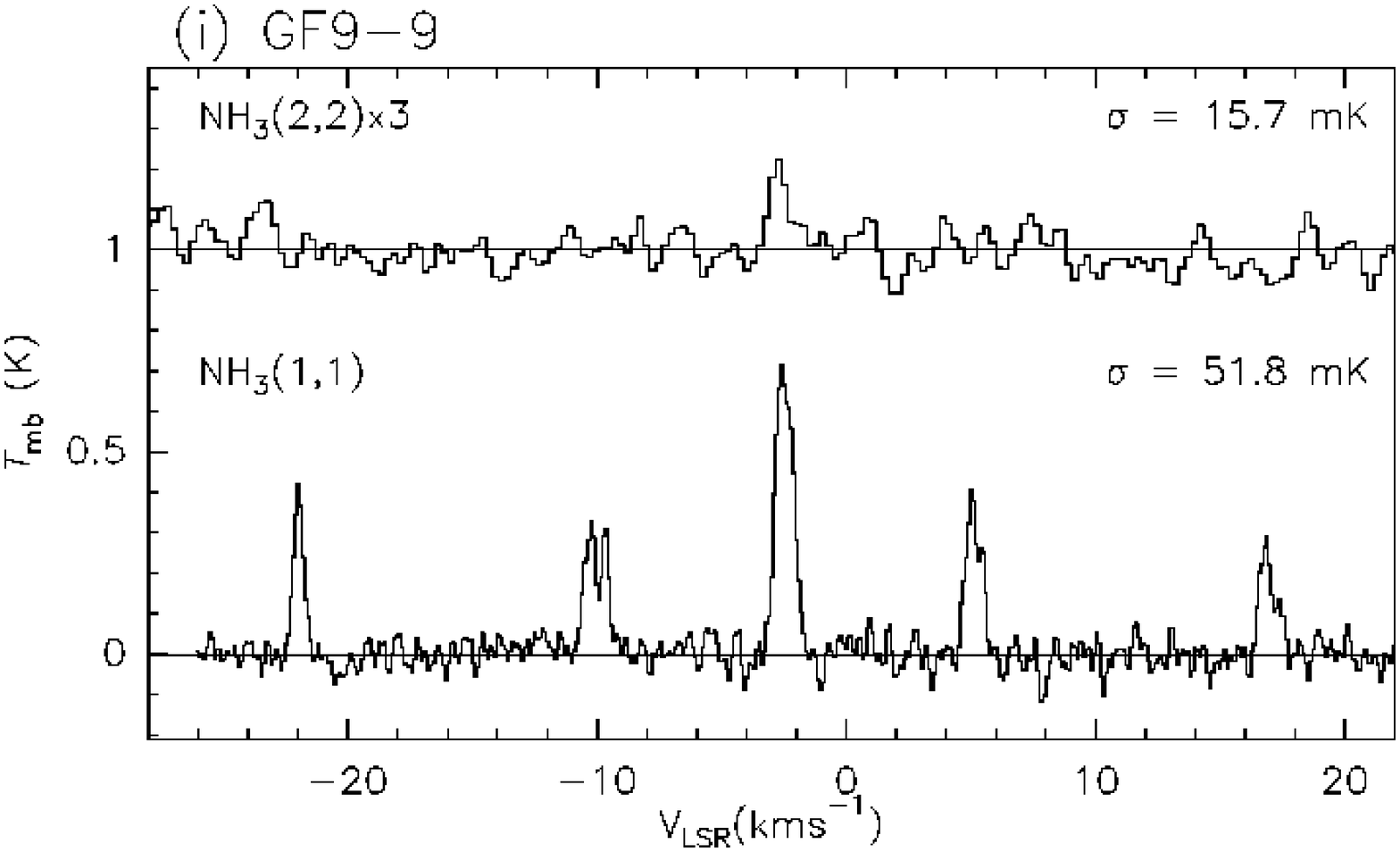}
\caption{Spectral profiles of the NH$_3$ (1,1) and (2,2) rotational 
inversion lines at the peak positions, except for GF\,9-6, 
of the identified cores and the candidates (see section \ref{ss:totmap}) 
obtained through the deep integration (section \ref{s:obs}).
Notice that the spectrum towards GF\,9-6 in panel (f) has not
been obtained at the center position of the core (see figure \ref{fig:eachmap}f),
and that no deep-integration has been carried out towards the GF\,9-10 core.
The $1\sigma$ RMS noise levels shown for each spectrum are given 
in \mbox{$T_{\rm mb}$} scale.
\label{fig:sp}}
\end{figure}

Figures \ref{fig:map} and \ref{fig:eachmap} show total integrated intensity maps of 
the NH$_3$ (1,1) emission for the whole filament and individual cores, respectively. 
Here, we integrated the main group of the hyperfine (HF) emission 
(Wilson, Bieging \& Downes 1978; UWW80)
between $V_{\rm LSR}$ $=-3.2$ km~s$^{-1}$ and $-1.4$ km~s$^{-1}$.
To find the velocity range, we made figure \ref{fig:vrange}
where we present the LSR-velocity ranges of the main HF groups for the 9 spectra 
(figure \ref{fig:sp}) taken with the deep integrations (section \ref{s:obs}).
The velocity range for each core is defined by the two LSR-velocities
where the intensity of the main HF group drops to the 1.5$\sigma$ level.
All 9 spectra showed similar velocity ranges;
the most blue- and redshifted velocities are found to be
$V_{\rm LSR}$ $=-3.21$ km~s$^{-1}$ in core 9 and 
$-1.36$ km~s$^{-1}$ in core 8, respectively.\par

Figure \ref{fig:map} clearly shows the presence of 7 dense cloud cores,
labeled GF\,9-5, 8, 4, 3, 2, 1, and 9 from the east to the west; 
table \ref{tbl:cores} summarizes the peak positions of the cores,
together with the other names found in the literature.
The peak \mbox{$T_{\rm mb}$} of these cores exceeds 
our detection threshold of S/N $=$ 6,
which means S/N$\geq$\,3 for the 50\% level contour with respect to the
peak intensity.
Notice that each core is detected at only a few observing points 
as shown in figure \ref{fig:eachmap}.
Figure \ref{fig:eachmap}f shows that the two core candidates GF\,9-6 and 10 have
peak intensities of 3 $\leq$ S/N $<$ 6.
They are probably real objects because 
they seem to contain IRAS sources within the 3$\sigma$ level contours. 
The GF\,9-10 core may be a new detection, although
we have to verify its presence by obtaining a better S/N.
Although the GF\,9-7 core candidate can be marginally recognized above the 3$\sigma$ 
level in figure \ref{fig:eachmap}g,
we clearly detected the (1,1) emission through the deep integration
(figure \ref{fig:sp}g).
The GF\,9-2, 3, and 4 cores were detected in NH$_3$ (1,1) with 
the Haystack 37\,m telescope 
($\theta_{\rm HPBW}$ $ \sim$ \timeform{87''}; BM89) 
and were designated as L\,1082C, A, and B, respectively.
The GF\,9-8 core is also seen in figure 33 of BM89, but these authors have not
given the core an identification number.
The GF\,9-8 core has not been observed in NH$_3$ mapping with the Effelsberg 100\,m telescope 
($\theta_{\rm HPBW}$ $=$ \timeform{40''} ; Wiesemeyer 1997).
It is interesting that the GF\,9-5 and 8 cores as well as the
GF\,9-7 core candidate do not exhibit YSO activity, suggestive of starless cores.\par

No dense core was detected towards the ``embedded YSO'' of LM\,349 (LM99),
while the two ``embedded YSOs'' of LM\,350 and 351 are associated with the NH$_3$ cores. 
Here, an ``embedded YSO'' is far-infrared bright YSO 
selected from the {\it IRAS} point source catalog (see LM99 for defintion).
The relationship between the remaining ``embedded YSO'' of LM\,352 and the GF\,9-4
core is not clear (see  figure \ref{fig:eachmap}d).
The absence of an NH$_3$ core around LM\,349 implies that 
the ``embedded YSO'' may be a very low-mass object whose
core mass is too small to be detected by our mapping survey.
We also point out that LM\,349 is close to 
one of the local peaks in the $^{13}$CO (1--0) column density map of Ciardi et al. 
(2000; see also figure 1 of Poidevin \& Bastien 2006).\par

It is worth noting that the IRAS sources located in the GF\,9-3 and -4
cores are driving weak molecular outflows with momentum rates of \mbox{$F_{\rm CO}$}, 
$\sim \,10^{-5}-10^{-6}$ \MO\ km~s$^{-1}$ yr$^{-1}$ (Bontemps et al. 1996),
approximately along the north-south direction.\par

Last, it is likely that 
the number density of the cores tends to become high towards the east, 
while low to the west. 
In fact, the four cores in the eastern part of the filament, 
GF\,9-3, 4, 5, and 8 seem to be confined to a rather small region of $\sim$0.8\,pc.
Since the mean separation between the two neighboring cores is 0.36 pc,
they can be treated as a group of cores.
Similarly, we consider the GF\,9-6 and -10 core candidates as another group.
Here, the separations are calculated using the peak positions in table \ref{tbl:cores}.
Consequently, the two core groups, the isolated GF\,9-2, 1, and 9 cores,
and the candidate core GF\,9-7 are located
at regular intervals of $\sim$\,0.9\,pc.

%%%%%%%%%%%%%%%%%%%%%%%%%%%%%%%%%%%%%%%%%%%%%%%%%%%%%%%%%
\subsection{NH$_3$ Spectra towards the Cores}
\label{ss:sp}

Figure \ref{fig:sp} presents the 9 spectra of the NH$_3$ $(J, K) = (1,1)$ and (2,2) 
rotation inversion lines obtained by the deep integrations (section \ref{s:obs}).
All the observed positions showed the intense (1,1) emission with the distinct five groups 
of the HF components,
except for GF\,9-7, where the inner satellite HF groups at 
$V_{\rm LSR}$ $\sim$ +6 km~s$^{-1}$ and $-10$ km~s$^{-1}$ are barely recognized.
Notice that the NH$_3$ spectra for the core candidates GF\,9-6 and 7 were not taken
towards the exact peak positions of the cores
(see figure \ref{fig:eachmap}f and g).\par

In contrast to the (1,1) lines, 
sole the main HF group of the (2,2) transition was detected
towards only GF\,9-2 and -9 with S/N $\geq$\,3. 
Given the attained S/N in our observations,
non-detection of the satellite (2,2) HF groups does not give a stringent limit 
in the optical depth, namely, $\tau_{22} \leq 7$
since the intrinsic ratio of the main to satellite groups is about 15.9
for the (2,2) transition (Wilson et al. 1978).
Nevertheless, we believe that the (2,2) lines are optically thin because the observed
\mbox{$T_{\rm mb}$} is considerably lower than 
the excitation temperature of the (1,1) transition,
\mbox{$T_{\rm ex}$} (1,1) (7.4 -- 9.5\,K; described in section \ref{ss:Ncol}).\par

%%%%%%%%%%%%%%%%%%%%%%%%%%%%%%%%%%%%%%%%%%%%%%%%%%%%%%%%%%%
\subsection{Hyperfine Structure Analysis and Column Density Calculations}
\label{ss:Ncol}

It is difficult to accurately know how each core is extended because
our survey observations have been conducted with full-beam sampling. 
We therefore limit our analysis to the peak spectra which have sufficient S/N,
instead of making source-averaged spectra.
To calculate beam-averaged NH$_3$ column densities (\mbox{$N_{\rm NH_3}$}),
we employed hyperfine structure (HFS) analysis,
which gives the total optical depth ($\tau_{\rm tot}$), 
the velocity width (\mbox{$\Delta v_{\rm FWHM}$}), 
and the LSR-velocity of one of the HF components ($v_0$).
Here, \mbox{$\Delta v_{\rm FWHM}$} is the FWHM of a single HF component and is assumed to be 
identical for all the HF components,
and $\tau_{\rm tot}$ is defined by the sum of the optical depths of all the HF components.
The \mbox{$\Delta v_{\rm FWHM}$} is subsequently deconvolved with the instrumental velocity resolution 
(section \ref{s:obs}) to estimate an intrinsic velocity width (\mbox{$\Delta v_{\rm int}$}).
It should be noted that \mbox{$\Delta v_{\rm int}$} is free from velocity width increase
caused by high optical depth because our HFS analysis solves \mbox{$\Delta v_{\rm FWHM}$} and
$\tau_{\rm tot}$ at the same time. We followed the procedure used in paper I;
our analysis is essentially equivalent to that summarized in 
Stutzki \& Winnewisser (1985) and other papers of, 
e.g., Winnewisser, Churchwell \& Walmsley (1979), 
UWW80, Pauls et al. (1983), and 
Ungerechts, Winnewisser \& Walmsley (1986).\par

First, we fitted one Gaussian profile to the main HF group of the 
(1,1) and (2,2) spectra. 
Table \ref{tbl:sp} summarizes the obtained peak temperatures in \mbox{$T_{\rm mb}$}, 
which are used to calculate the rotational temperature between 
the (1,1) and (2,2) levels (\mbox{$T_{\rm r,21}$}) in the third step.
Second, we performed the HFS analysis of the (1,1) spectra.
The resultant parameters of $v_0$, \mbox{$\Delta v_{\rm int}$}, and $\tau_{\rm tot}$ 
are summarized in table \ref{tbl:sp};
these $v_0$ values are plotted in figure \ref{fig:vrange} as well.
Third, using eq.(4) of Ho \& Towns (1983), 
we calculated \mbox{$T_{\rm r,21}$} with the above peak \mbox{$T_{\rm mb}$} values 
for both the transitions and $\tau_{\rm tot}$.
To use the equation, we gave the mean optical depth of the (1,1) line 
by $\tau_{\rm tot}$/18, where the denominator of 18 is the 
number of (1,1) HF components (e.g., UWW80).
For the 7 cores where the (2,2) transition was not detected, 
we adopted the 3$\sigma$ upper limit as a peak \mbox{$T_{\rm mb}$}. 
Table \ref{tbl:sp} shows that \mbox{$T_{\rm r,21}$} 
was calculated to be 7.4$\pm$0.3\,K for GF\,9-2 and
7.9$\pm$0.4\,K for GF\,9-9, 
while the other cores have upper limits ranging between 7.1\,K and 12.8\,K.
Fourth, given $\tau_{\rm tot}$, \mbox{$\Delta v_{\rm int}$}, 
and excitation temperature (\mbox{$T_{\rm ex}$}), 
we calculated the beam-averaged column density of NH$_3$
molecules in the (1,1) level, $N_{11}$, 
leading to \mbox{$N_{\rm NH_3}$} with the \mbox{$T_{\rm r,21}$} value.
For this purpose, we used eq.(2) in UWW80.
We assumed that all the energy levels are in LTE at temperature \mbox{$T_{\rm r,21}$},
and that \mbox{$T_{\rm r,21}$} is equal to \mbox{$T_{\rm ex}$} as well as the
gas kinetic temperature (\mbox{$T_{\rm k}$}) of the core, i.e., 
\mbox{$T_{\rm r,21}$} $=$ \mbox{$T_{\rm ex}$} $=$ \mbox{$T_{\rm k}$}.
The cores are dense ($n_{\rm H_2}\sim 10^5$ cm$^{-3}$) enough
to make our assumption valid (Stutzki \& Winnewisser 1985; Taffala et al. 2004).
We believe that this assumption is reasonable because
the derived \mbox{$T_{\rm r,21}$} agrees well with \mbox{$T_{\rm ex}$} of C$^{18}$O (1--0) 
(7--8\,K for GF\,9-2, 3, and 4; Myers, Linke \& Benson 1983),
$^{13}$CO (1--0) (7.2\,K for GF\,9-2; Ciardi et al. 2000), 
and N$_2$H$^+$\ (1--0) (9.5\,K; paper I).
We consequently obtained the NH$_3$ column densities
as summarized in table \ref{tbl:property}.
Here, we exclude the candidate cores No.\,6, 7, and 10 with insufficient S/N.
We found that the error in $\tau_{\rm tot}$ 
and the possible uncertainty in \mbox{$T_{\rm ex}$}
of 7.4 -- 9.5\,K caused uncertainties in $N_{11}$ 
ranging from $3\%$ (GF\,9-2) to 33\% (GF\,9-5).\par

Last, we verified whether or not the above results towards the GF\,9-2 core are
consistent with those obtained from the spectra with coarse velocity resolution
in paper I (see the core center portions of tables 5 and 6).
It should be noted that the optical depth given in table 5 of paper I is 
that for the most intense HF component, which can be converted into 
$\tau_{\rm tot}$ of 6.8$\pm$0.4
and the value does not significantly 
differ from the new value of 8.0$\pm$0.2 in this work.
The high velocity resolution in this study gives $\sim 1.5$ times narrower
\mbox{$\Delta v_{\rm int}$} than that in the previous work. 
The most significant difference between the two studies is that
the new observations have succeeded in detecting the (2,2) transition, 
leading to the low \mbox{$T_{\rm r,21}$} values.
In other words, the cold temperature is the main cause of the higher $N$(NH$_3$), 
by a factor of $\sim$\,3 compared with paper I.

%%%%%%%%%%%%%%%%%%%%%%%%%%%%%%%%%%%%%%%%%%%%%%%%%%%%%%%%%
\subsection{LTE Masses of the Cores}
\label{ss:MLTE}

Once the beam-averaged \mbox{$N_{\rm NH_3}$} has been obtained, one can calculate the LTE-mass, 
\mbox{$M_{\rm LTE}$}, within the beam by supplying the fractional abundance of NH$_3$ 
molecules, $X$(NH$_3$).
We assumed that $X$(NH$_3$) $=$ \mbox{$(2\pm 1)\times 10^{-8}$}, estimated in GF\,9-2 (paper I), 
is valid for the other cores. 
Notice that, as addressed in paper I,
the adopted abundance has reasonable agreement with
those in similar objects (Jijina, Myers \& Adams 1999).
The resultant \mbox{$M_{\rm LTE}$}\ ranges from 
1.8 \MO\ (GF\,9-5) to 8.2 \MO\ (GF\,9-2)
(table \ref{tbl:property}).
Although the LTE-mass for GF\,9-2 seems to be twice the previous estimate of 4.5$\pm$2.4 \MO,
the two measurements can be reconciled considering
the $\sim$\,50\% uncertainty in $X$(NH$_3$ ) and the difference between the 
adopted core sizes (5.5$\times 10^{-3}$ pc$^2$ in paper I, 
versus 3.9$\times 10^{-3}$ pc$^2$ in this work).
Notice that the $\sim$\,50\% uncertainty in $X$(NH$_3$) causes
a significantly larger error in \mbox{$M_{\rm LTE}$} than those
from the uncertainties in $\tau_{\rm tot}$ and \mbox{$T_{\rm ex}$}
described in the previous subsection.\par

%%%%%%%%%%%%%%%%%%%%%%%%%%%%%%%%%%%%%%%%%%%%%%%%%%%%%%%%%%%
\subsection{Velocity Widths and Virial Masses of the Cores}
\label{ss:dV}

In table \ref{tbl:sp}, we summarize the \mbox{$\Delta v_{\rm int}$} 
values obtained through the HFS analysis.
Clearly, the observed intrinsic velocity widths in all the cores are
2--3 times larger than \mbox{$\Delta v_{\rm thm}$} 
of 0.16 -- 0.18 km~s$^{-1}$ for the \mbox{$T_{\rm k}$} range of 7.4 -- 9.5 \,K
[\mbox{$\Delta v_{\rm thm}$} $=\{8(\ln 2)kT_{\rm k}/m_{\rm NH_3}\}^{1/2}$].
This fact indicates that the internal motions of all the cores are dominated by 
non-thermal pressure (\mbox{$\Delta v_{\rm nth}$}), such as supersonic turbulence.
Assuming that the non-thermal motions as well as the thermal ones
support the cores against their self-gravity, 
we calculated  virial masses (\mbox{$M_{\rm vir}$} ; see table \ref{tbl:property})
using \mbox{$\Delta v_{\rm int}$} in table \ref{tbl:sp}.
In section \ref{ss:virial}, we compare \mbox{$M_{\rm LTE}$} with \mbox{$M_{\rm vir}$}\ 
in the context of the dynamical instability of the cores.

%%%%%%%%%%%%%%%%%%%%%%%%%%%%%%%%%%%%%%%%%%%%%%%%%%%%%%%%%
\section{Discussion}
\label{s:Discussion}

\subsection{Gravitational Instability of the Cores}
\label{ss:virial}

It is probable that all the cores in the GF\,9 filament are gravitationally unstable 
because \mbox{$M_{\rm vir}$}\ $<$ \mbox{$M_{\rm LTE}$} (table \ref{tbl:property}) 
if we interpret that 
the non-thermal velocity widths are due to the supersonic 
turbulence within the cores. 
Here, another distance estimate of 440\,pc (section \ref{s:intro}) does not alter the
relationship of \mbox{$M_{\rm vir}$}\ $<$ \mbox{$M_{\rm LTE}$} 
because \mbox{$M_{\rm vir}$}\ is proportional to $d$ while
\mbox{$M_{\rm LTE}$} varies with $d^2$.
To verify the interpretation, we should assess origin of the non-thermal widths, 
as done in Ciardi et al. (2000) and paper I. 
In addition to the turbulent motions, 
it is well known that spectral line profiles can be broadened beyond their thermal widths 
owing to systematic motions of gas such as stellar winds or/and infall.
However, it is unlikely that \mbox{$\Delta v_{\rm nth}$} arises
primarily from supersonic stellar winds or outflows because, 
to our knowledge, the NH$_3$ inversion lines do not trace molecular outflows,
but static dense gas.\par

We suggest that the non-thermal velocity widths of the cores are produced 
by the large-scale infall motions rather than the turbulent ones. 
In fact, we discussed the presence of the supersonic infalling 
motions all over the GF\,9-2 core
in paper I. Our recent follow-up observations have clearly detected blue-skewed profiles in 
the optically thick HCO$^+$ (3--2) and (1--0) lines over the
GF\,9-2 core (R. S. Furuya et al., in prep.).
Moreover, we have detected such blue-skewed profiles towards
the peak positions of the GF\,9-4, 8, and 9 cores with the (3--2) transition.
These facts reinforce the idea that 
the internal motions of the cores in the GF\,9 filament are 
dominated by the large-scale supersonic 
infalling motions, although we have to assess the presence of the infall
in all the cores through mapping observations.
If this interpretation is valid for all the cores,
\mbox{$M_{\rm vir}$}\ should be calculated solely with the thermal width, 
i.e., \mbox{$M_{\rm vir, thm}$}
$= \frac{5}{32\ln{2}}\cdot\frac{\theta_{\rm HPBW}}{G}\cdot\frac{kT_{\rm k}}{\mu m_{\rm H}}$, 
yielding 0.19 -- 0.25 \MO\ for the \mbox{$T_{\rm ex}$} range of 7.4 -- 9.5\,K.
This estimate suggests that most of the cores should be gravitationally unstable 
because \mbox{$M_{\rm vir, thm}$} $<$ \mbox{$M_{\rm LTE}$} 
in spite of the $\sim 50\%$ uncertainty in \mbox{$M_{\rm LTE}$}.\par

\subsection{Formation of the Cores in the Filament}
\label{ss:filament}

A gravitational fragmentation process can explain why the dense cores 
in the GF\,9 filament are located at regular intervals of $\sim$0.9\,pc 
(see figure \ref{fig:map}).
Recall that we consider the GF\,9-5, 8, 4, and 3 cores, as well as the GF 9-6 and 10 core candidates,
as groups ($\S\ref{ss:totmap})$.
For simplicity, we compare the core separation with an expected Jeans length
($\lambda_{\rm J}$) for the filament.
Ciardi et al. (2000) estimated a density of $n_{\rm H_2} = \,1700\pm$200 cm$^{-3}$ 
for the GF\,9 filament from their $^{13}$CO (1--0) observations.
Assuming that the ambient gas has a typical density of $\sim 10^3$ cm$^{-3}$, 
$\lambda_{\rm J}$ is computed to be $\sim$\,0.6\,pc at \mbox{$T_{\rm k}$} $=$ 10\,K.
This estimate is not significantly different from the core separation of $\sim$\,0.9 pc. 
Here, the temperature of 10\,K is taken from the typical 
peak \mbox{$T_{\rm mb}$} (6--7\,K) of the optically thick
$^{12}$CO (3--2) and (1--0) emission derived in the ambient gas around the GF\,9-2 core (paper I).
In addition, the Jeans mass for \mbox{$T_{\rm k}$} $=$ 10\,K and $n_{\rm H_2}\sim 10^3$ cm$^{-3}$ is
calculated to be $\sim 3$ \MO\ which is comparable to the core LTE-masses
(table \ref{tbl:property}).
Such an estimate based on the "classical" Jeans analysis implies that 
the fragmentation of the filament has been caused by the gravitational instability.\par

Alternatively, the spatial distribution of the identified cores might be understood 
in terms of the magnetohydrodynamical instability of the GF\,9 filamentary cloud,
as Hanawa et al. (1993) proposed for interpreting the ongoing fragmentation process in the
Orion A filamentary cloud. They discussed that 
the effective sound velocity increases when the effect of
magnetic field and/or rotation is considered [see their eq.(11)],
and that the Jeans length of the filament becomes as short as the core separation
unless the filament is almost perpendicular to the plane-of-sky. 
The apparent configuration of the GF\,9 filament would be somewhat
similar to that of the Orion cloud.
However, to make such a comparison, 
we need to know not only basic parameters characterizing the filament
such as length, width, inclination angle, and total mass, but also
magnetic field strength and/or velocity field.

\section{Summary}
Using the Nobeyama 45\,m telescope, we carried out an unbiased
survey of dense molecular cores in the GF\,9 filament.
The obtained large-scale map of the NH$_3$ (1,1) emission revealed that the filament contains
7 dense ($n_{\rm H_2}\sim 10^4$ cm$^{-3}$) and cold (\mbox{$T_{\rm k}$} $\lesssim$ 10\,K) cores 
having the LTE-masses of 1.8 -- 8.2 \MO\ and the three candidates located at the
regular intervals of $\sim 0.9$\,pc.
We argued that these cores appear to be gravitationally unstable and have formed 
through the gravitational fragmentation of the natal filamentary cloud.
Further high-resolution imaging of the identified cores, 
as well as a search for blue-skewed infall profiles 
over the cores will allow us to discuss the physical properties of the cores 
on a more solid ground.
We lastly point out that good knowledge of the velocity field of the low-density inter-core gas 
traced by, e.g., the $^{13}$CO (1--0) line
will provide us with an essential clue towards understanding the formation mechanism of
star forming cores in a filamentary cloud.

\bigskip

The authors sincerely thank Thomas A. Bell for critical reading of the paper.
R. S. F. thanks Isabel de Maurissens for providing a copy of
the Viotti (1969) paper which is published in 
Memorie della Societa Astronomia Italiana. 
It is a great pleasure to thank the staff of NRO for their generous 
help during the observations.
This work is partially supported by a Grant-in-Aid for
Scientific Research (A) from the Ministry of Education,
Culture, Sports, Science and Technology of Japan
(No.\,19204020). 

%%%%%%%%%%%%%%%%%%%%%%%%%%%%%%%%%%%%%%%%%%%%%%%%%%%%%%%%%

\newpage

\newpage

\newpage

\newpage

\begin{table}[htbp]
\rotatebox{90}{
\begin{minipage}{\textheight}
\begin{center}
\caption{Peak Positions of the NH$_3$ Cores including the Candidates in the GF\,9 Filament}
\label{tbl:cores}
\begin{tabular}{lllllll}
\hline\hline
\lw{GF\,9-$^*$} & \multicolumn{2}{c}{Other Identifications$^{\dagger}$} & & \multicolumn{2}{c}{Peak Position$^\ddagger$} & Associated \\
\cline{2-3}\cline{5-6}
                    & W97      & BM89       & & R.A.(J2000.0) & Dec.(J2000.0) & YSO \\
\hline
5 core              & GF\,9-5        & $\cdot\cdot\cdot$    & & \timeform{20h54m01s} & \timeform{60D24'09"} & $\cdot\cdot\cdot$ \\
3 core              & GF\,9-3A/3B    & L\,1082A    & & \timeform{20h53m25s} & \timeform{60D14'11"} & IRAS\,20520+6003 \\
4 core              & GF\,9-4        & L\,1082B    & & \timeform{20h53m49s} & \timeform{60D09'30"} & IRAS\,20526+5958, LM\,352 \\
8 core              & $\cdot\cdot\cdot$ & $\cdot\cdot\cdot$    & & \timeform{20h53m52s} & \timeform{60D17'10"} & $\cdot\cdot\cdot$ \\
2 core$^{\S}$ & GF\,9-2        & L\,1082C    & & \timeform{20h51m29s} & \timeform{60D18'34"} & PSC\,20503+6006, LM\,351 \\
1 core              & GF\,9-1        & $\cdot\cdot\cdot$    & & \timeform{20h49m50s} & \timeform{60D15'52"} & LM\,350 \\
9 core              & GF\,9-i6       & $\cdot\cdot\cdot$    & & \timeform{20h47m57s} & \timeform{60D03'24"} & IRAS\,20468+5953 \\
6                   & GF\,9-6        & $\cdot\cdot\cdot$    & & \timeform{20h47m05s} & \timeform{59D51'18"} & IRAS\,20460+5939 \\
10                  & $\cdot\cdot\cdot$ & $\cdot\cdot\cdot$    & & \timeform{20h46m28s} & \timeform{59D48'53"} & IRAS\,20454+5938 \\
7		    & GF\,9-7        & $\cdot\cdot\cdot$     & & \timeform{20h48m11s} & \timeform{59D39'25"} & $\cdot\cdot\cdot$ \\
\hline
\end{tabular}
\end{center}
%%%%%%%%%%%%%%%%%%%%%%%%%%%%%%%%%%%%%%%%%%%%%%%%%%
%%%%%%%%%%%%%%%%%%%%%%%%%%%%%%%%%%%%%%%%%%%%%%%%%
\begin{flushleft}
\footnotemark[$*$] GF\,9 core designation numbers with the label of 
``core'' are for the firm detection, while the others for the possible detections.
See sections \ref{ss:totmap} and \ref{ss:Ncol} for details.
We used the same designation number for the cores used in Wiesemeyer et al. (1997) 
to keep consistency as much as possible.
Notice that the above core numbers are different from the GF\,9 star designation
numbers with upper case S given in Poidevin \& Bastien (2006). \\
\footnotemark[$\dagger$] References --- W97:  Wiesemeyer et al. (1997), BM89: Benson \& Myers (1989) \\
\footnotemark[$\ddagger$] The errors in the absolute positions, due to the telescope pointing error
(see section \ref{s:obs}), are typically \timeform{5"}. \\
\footnotemark[$\S$] GF\,9-2 core is also identified as GF\,9-Core in Ciardi et al. (2000). \\
\end{flushleft}
%%%%%%%%%%%%%%%%%%%%%%%%%%%%%%%%%%%%%%%%%%%%%%%%%%
\end{minipage}
}
\end{table}

\newpage
\begin{table}[htbp]
\rotatebox{90}{
\begin{minipage}{\textheight}
\begin{center}
\caption{Summary of the NH$_3$ Spectral Line Analysis$^\ast$}
\label{tbl:sp}
%%%%%%%%%%%%%%%%%%%%%%%%%%%%%%%%%%%%%%%%%%%%
\begin{tabular}{cccccccccr}
\hline
\lw{} & \multicolumn{5}{c}{NH$_3$ (1,1)} & & \multicolumn{1}{c}{NH$_3$ (2,2)} & & \\
\cline{2-6}
\cline{8-8}
Core$^\dagger$ & $T_{\rm mb}^\ddagger$ & & $v_0^\S$ & $\Delta v_{\rm int}^\S$ & \lw{$\tau_{\rm tot}^\S$} & & $T_{\rm mb}^\ddagger$ & & $T_{\rm r,21}^\P$ \\
     & (mK) &  &          (km~s$^{-1}$) &          (km~s$^{-1}$) &              & & (mK) && (K) \\
\hline\hline
1 &  860$\pm$50 &  & $-2.53$ & 0.37$\pm$0.02 & 4.4$\pm$0.5  & & $\leq 72$ & & $\leq$  7.1 \\
2 & 1270$\pm$90 &  & $-2.52$ & 0.39$\pm$0.01 & 8.0$\pm$0.2  & & 110$\pm$9 & & 7.4$\pm$0.3 \\
3 &  740$\pm$55 &  & $-2.23$ & 0.42$\pm$0.02 & 6.0$\pm$0.7  & & $\leq 69$ & & $\leq$  7.6 \\
4 &  630$\pm$40 &  & $-2.36$ & 0.31$\pm$0.02 & 6.9$\pm$0.9  & & $\leq 62$ & & $\leq$  7.8 \\
5 &  320$\pm$25 &  & $-2.32$ & 0.39$\pm$0.03 & 1.8$\pm$0.6  & & $\leq 29$ & & $\leq$  9.7 \\
6 &  310$\pm$20 &  & $-2.42$ & 0.28$\pm$0.02 & 2.1$\pm$0.6  & & $\leq 29$ & & $\leq$ 10.1 \\
7 &  160$\pm$12 &  & $-2.14$ & 0.50$\pm$0.08 & 2.0$\pm$1.3  & & $\leq 40$ & & $\leq$ 12.8 \\
8 &  630$\pm$50 &  & $-2.15$ & 0.34$\pm$0.02 & 8.4$\pm$1.0  & & $\leq 87$ & & $\leq$  7.9 \\
9 &  600$\pm$45 &  & $-2.61$ & 0.43$\pm$0.02 & 6.3$\pm$0.6  & & 57$\pm$6 & & 7.9$\pm$0.4 \\
\hline
\end{tabular}
\end{center}
%%%%%%%%%%%%%%%%%%%%%%%%%%%%%%%%%%%%%%%%%%%%%%%%%%
\begin{flushleft}
\footnotemark[$\ast$] We analyzed the peak spectra shown in figure \ref{fig:sp} (see section \ref{ss:Ncol} for the details).\\
\footnotemark[$\dagger$] All the spectra, except for the core candidates of \,\#6 and \#7,
have been taken towards the center positions of the identified cores (see figure \ref{fig:eachmap}).\\
\footnotemark[$\ddagger$] Peak \mbox{$T_{\rm mb}$} for the main group of the hyperfine emission obtained by single Gaussian fitting. The upper limits for the (2,2) transition are the 3$\sigma$ upper limits 
(see figure \ref{fig:sp} as well). \\
\footnotemark[$\S$] $v_0$, \mbox{$\Delta v_{\rm int}$}, and $\tau_{\rm tot}$ denote, respectively, the centroid velocity,
the intrinsic velocity width after correcting for the instrumental velocity resolution, and 
the total optical depth obtained from the hyperfine structure analysis
(see section \ref{ss:Ncol} for the details).
The errors for $v_0$ are typically less than 0.01 km~s$^{-1}$.\\
\footnotemark[$\P$] Rotational temperature of NH$_3$ molecules 
between the (1,1) and (2,2) states (see section \ref{ss:Ncol}.).\\
\end{flushleft}
%%%%%%%%%%%%%%%%%%%%%%%%%%%%%%%%%%%%%%%%%%%%%%%%%%
\end{minipage}
}
\end{table}

\newpage
\begin{table}[htbp]
\rotatebox{90}{
\begin{minipage}{\textheight}
\begin{center}
\caption{Properties of the Ammonia Cores in the GF\,9 Filament
\label{tbl:property}}
\begin{tabular}{ccccc}
\hline\hline
%%%%%%%%%%%%%%%%%%%%%%
\lw{Core} & $M_{\rm vir}^\ast$ & $N({\rm NH_3})^\dagger$    & $M_{\rm LTE}^\ddagger$ \\
                & (\MO )                      & ($\times 10^{15}$ cm$^{-2}$) &  (\MO) \\
%%%%%%%%%%%%%%%
\hline
%%%%%%%%%%%%%%%
1 & 1.0$\pm$0.09 & 1.0$\pm$0.2  & 4.3$\pm$2.1 \\
2 & 1.1$\pm$0.03 & 2.0$\pm$0.3  & 8.2$\pm$4.1 \\
3 & 1.3$\pm$0.16 & 1.6$\pm$0.3  & 6.7$\pm$3.3 \\
4 & 0.7$\pm$0.08 & 1.4$\pm$0.3  & 5.7$\pm$2.9 \\
5 & 1.1$\pm$0.18 & 0.44$\pm$0.1 & 1.8$\pm$0.9 \\
8 & 0.9$\pm$0.09 & 1.8$\pm$0.3  & 7.5$\pm$3.8 \\
9 & 1.3$\pm$0.12 & 1.7$\pm$0.4  & 7.1$\pm$3.6 \\
%%%%%%%%%%%%%%%
\hline
\end{tabular}
\end{center}
%%%%%%%%%%%%%%%%%%%%%%%%%%%%%%%%%%%%%%%%%%%%%%%%%%
%%%%%%%%%%%%%%%%%%%%%%%%%%%%%%%%%%%%%%%%%%%%%%%%%%
\begin{flushleft}
\footnotemark[$\ast$] Virial mass of the cores calculated from 
\mbox{$M_{\rm vir}$}\ 
$=\frac{5}{16\ln{2}}\cdot \frac{\theta_{\rm HPBW}}{G}\cdot \left(\Delta v_{\rm int}\right)^2$ 
(see section \ref{ss:dV}) where $\theta_{\rm HPBW}$ is \timeform{73''}.\\
\footnotemark[$\dagger$] Beam-averaged NH$_3$ column density; the errors are calculated from the possible \mbox{$T_{\rm ex}$} range of 7.4 -- 9.5\,K and the uncertainties in $\tau_{\rm tot}$ and \mbox{$\Delta v_{\rm int}$}.\\
\footnotemark[$\ddagger$] LTE-mass of the cores calculated from \mbox{$M_{\rm LTE}$} $=\mu_{\rm g}m_{\rm H}\cdot \frac{N(\rm NH_3)}{X(\rm NH_3)}\cdot\pi\left(\frac{\theta_{\rm HPBW}}{2}\right)^2$ (see section \ref{ss:MLTE}). 
Here, $\mu_g$ denotes the mean molecular weight of 2.33 and $m_{\rm H}$ the atomic mass of hydrogen. The errors are due to the $\sim$\,50\, \% uncertainty in the $X(\rm NH_3)$. \\
\end{flushleft}
%%%%%%%%%%%%%%%%%%%%%%%%%%%%%%%%%%%%%%%%%%%%%%%%%%
\end{minipage}
}
\end{table}
%%%%%%%%%%%%%%%%%%%%%%%%%%%%%%%%%%%%%%%%%%%%%%%%%%%%%%%%%%%%%%%%%%%%%%%

\end{document}